\documentclass[12pt]{article}
\usepackage{booktabs}
\usepackage{amsthm}
\usepackage{amsmath}
\usepackage{mathrsfs}
\usepackage{graphicx}
\usepackage{authblk}
\usepackage{times}

\usepackage{pdflscape}
\usepackage{float}
\usepackage[authoryear,sort&compress,round]{natbib}

\usepackage{hyperref}
\hypersetup{
    colorlinks=true,
    linkcolor=red,
    filecolor=blue,
    urlcolor=blue,
    citecolor=blue,
}
\usepackage{geometry} 
\usepackage[normal]{caption2}

\usepackage{indentfirst}
\begin{document}
\date{}
 \title {Substantial but heterogeneous impacts of high-speed rail on talent flow in China}
 \author[a]{Mei Xie}
 \author[b]{Jian Gao \footnote{Corresponding author: jian.gao1@kellogg.northwestern.edu}}
 \author[a]{Tao Zhou \footnote{Corresponding author: zhutou@ustc.edu.}}

 \affil[a]{Big Data Research Center, University of Electronic Science and Technology of China, Chengdu 611731, China}
 \affil[b]{Kellogg School of Management, Northwestern University, Evanston, IL 60208, USA}
\maketitle
\begin{abstract}
The great expansion of high-speed rail (HSR) in China facilitates communications and interactions among people across cities. Despite extensive literature documenting the effects of HSR on a variety of variables such as local economic development, research collaboration, tourism, and capital mobility, not much is known about how HSR affects the flow of well-educated workers, says talents. Here we estimate talent flow among Chinese cities based on large-scale resume data of over 3.4 million online job seekers and explore how it is affected by HSR. Specifically, we employ both a multiple linear regression model that controls for several socioeconomic factors and a two-stage least square regression model that instruments the introduction of HSR to a city to address endogeneity concerns. We find that the introduction of HSR has an overall positive effect on the talent net inflow of a city although both inflow and outflow are increased. Moreover, the effects of HSR on talent flow are rather heterogeneous for cities with different levels of economic development and for talents working in different industries. Specifically, developed cities benefit from HSR, whereas less-developed cities are relatively impaired. Cities connected by HSR show significant advantage in attracting talents from secondary and tertiary industries. These substantial but heterogeneous effects of HSR suggest a desire for more comprehensive thinking about the long-term benefits of entering the HSR network, especially for less-developed cities and those with comparative advantage in manufacturing and service industries.

\textbf{Keywords:} High-speed rail, Talent flow, Regional development, Heterogeneous analysis, Moderating effects
\end{abstract}

\section{Introduction}

    Transportation infrastructures, such as highways, railways, and airports, are fundamental linkages of intercity economic activities as they facilitate the flow of people and goods within and across geographic boundaries \citep{li2020estimation, jiao2020roles}. Among these infrastructures, high-speed rail (HSR) is a safe, comfortable, and environmentally-friendly transport tool \citep{lingreenhouse}, which plays a vital role in increasing economic mobility by reducing travel time \citep{vickerman2006indirect,chen2012wider}. Since the Japanese Shinkansen opened in 1964, many countries and regions such as Germany, Spain, and France have invested in HSR projects. According to the National Railway Administration of China, HSR lines refer to passenger transport railway with a designed speed of at least 250 km/h and an operation speed of at least 200 km/h \citep{lawrence2019china}. In August 2008, China's first passenger dedicated HSR line (the Beijing-Tianjin intercity rail) operated at a maximum speed of 350 km/h \citep{amos2010high}. At the end of 2019, the total mileage of China's HSR lines exceeded 35,000 km.

    HSR helps people overcome the distance barrier so that labors have more opportunities to pursue high levels of salaries and living environment \citep{jiao2017impacts,yang2018study,zhou2018implications,fan2019connectivity}. By facilitating passenger mobility for entertainment and business, HSR accelerates the flow of knowledge, experience, information, and other economic factors \citep{wang2019impact,duan2020transportation}, and thus it intensifies economic activities between cities, such as enterprise investment \citep{lin2019facilitating}, regional cooperation \citep{wukang159}, innovation output \citep{dong2020role}, technological communication \citep{lintechnology}, and market transaction \citep{cascetta2011analysis,campa2016high,pagliara2017exploring}, eventually leading to better allocation of economic factors and faster growth of regional economy \citep{chen2011impacts,cheng2015high,long2018high, diao2018does, pan2020high, gao2021spillovers}. The reduction in travel time also boosts the development of transportation, tourism, retail trade, and hotel \citep{lin2017travel,gao2019does, dong2020role}. Moreover, HSR is conducive to enhance the capital flow and expand the scale of enterprises \citep{cui2019high,lin2019facilitating,yang2019does,zou2019high}, resulting in an increase in employment opportunities. HSR is expected to affect labor mobility and redistribution \citep{yin2015effects} for the improvement of accessibility and reduction of commuting time.

    Previous studies show multiple effects of HSR on labor flow. On the one hand, the improvement in accessibility can help in diffusing economic activities from core to periphery, leading to an expansion of investment, trade, and market of peripheral cities \citep{alonso1964location,muth1969cities}. As a result, HSR increases attractiveness of peripheral cities \citep{liang2020effectiveness}. On the other hand, HSR reduces the travel time between core and peripheral cities, which makes it easier for labors to commute between peripheral cities and developed cities \citep{helpman1985market,krugman1991increasing}. Under the siphoning effects \citep{qin2017no, ke2017china, yu2019highspeed, gao2020does,dong2021ghost}, both human capital and production resources migrate from less-developed to developed cities, and thus peripheral cities suffer from the economic downturn, declination in labor productivity, and population loss \citep{qin2017no,ke2017china,yang2019does}, resulting in enlarged economic unbalance between core and periphery.

     Empirical evidences have suggested correlations between HSR network and population flow. For example, based on hiring data in Spain during 2002-2014, \citet{guirao2018labour} analyzed the impact of HSR on inter-regional labor migration at the province level and found that both HSR services and the location of HSR significantly affect labor mobility. A majority of studies have found that HSR leads population to concentrate in megacities due to the siphoning effects \citep{yang2019does,gao2020does}. \citet{deng2019shrinking} found that HSR exacerbates population loss in shrinking cities during 2006-2015. \citet{xu2020siphon} explored the impact of HSR on internal migration and found that megacities gained more population inflowing. \citet{wang2019impact} found that HSR promotes short-term population mobility but has a negative influence on long-term population migration. \citet{kim2015impacts} found that after the opening of HSR between Seoul and Pusan, the population concentrated in Seoul and its fringe areas. \citet{sasaki1997high} suggested that the Shinkansen HSR could not effectively disperse the population from core to periphery. Other scientists document that the space-time compression effect of HSR will lead to population outflow from megacities, and thus will facilitate the population growth of medium and small-sized cities along the HSR line \citep{liang2020effectiveness, huang2021spatial}. \citet{heuermann2019effect} explored the quantitative relationship between reducing travel time and commuters by using the gravity model. They found that a 1\% decrease in travel time is associated with a 0.25\% increase in the number of commuters. \citet{dong2020role} showed that academic researchers prefer to move to second-tier cities with HSR stations.

    In despite of a large body of literature studying the impacts of HSR on population flow, studies on the flow of well-educated workers (hereafter talents) are rare, largely due to the lack of relevant large-scale and timely data. In the era of the knowledge economy, talents with professional knowledge and skills are the key to boost knowledge innovation, technological progress, and economic outputs. In other words, a city with a larger number of talents possesses a higher labor productivity. The fast expansion of HSR can accelerate the flow of talents to look for better jobs, leading to the reallocation of human capital among cities. Up to now, what role does HSR play in boosting talent flow between cities and whether the impacts of HSR on talent flow are heterogeneous are relatively underexplored. Answering these questions will not only guide the rest of the Chinese cities on whether to join the HSR network but also inform other countries on whether to build an analogous HSR network like China for boosting their economy.

    In this paper, we investigate the impacts of HSR on talent flow at the city level based on large-scale resume data from Chinese online job seekers. Here we treat job seekers who have obtained a bachelor’s degree or higher as talents \footnote{According to the China Labor Statistics Yearbook (2016), about 17\% of in total 775 million employed people in China have a bachelor’s degree or higher.}. The resume data tracks an individual’s career path and job expectation, from the city of current residence to the city of the expected job in 2015. Accordingly, we estimate the relative intensity of inflow, retention, outflow, and net inflow of talents for each city. We explore the effects of HSR on talent flow by a multiple regression model with controlling for several social and economic factors and by a two-stage least square regression model that instruments the introduction of HSR to a city. Cities with one or more HSR stations operating in 2014 are categorized into the treatment group and others make up the control group. We find that the opening of HSR boosts talent inflow by 53.9\%, talent retention by 48.5\%, and talent outflow by 39.9\%. HSR increases the net inflow by 14\%. We further address the underlying endogeneity of HSR station by using an instrument variable construed by the least cost HSR network. Results support the robustness of our observations that HSR has positive effects on talent flow. Moreover, we find that the impact of HSR on talent flow is rather heterogeneous. First, the economic status of cities moderates the relationship between HSR and talent flow such that the positive impact of HSR on talent net inflow is more pronounced for well-developed cities while HSR exacerbates the brain drain for less-developed cities. Second, HSR plays an important role in promoting the talent flow of secondary and tertiary industries.

    The rest of the paper is organized as follows: Section \ref{s2} presents the data, variables, and methods. Section \ref{s3} reports results. Section \ref{s5} summarizes the work and discusses the relevance of our findings.

\section{Data and methods}\label{s2}
\subsection{Data}
    In this study, we collected various datasets from different sources, including resume data, HSR data, and socioeconomic data.

    \textbf{Resume Data.} The talent flow is derived from anonymized resume data of Chinese online recruitment website, covering about 10.3 million individuals. The detailed information of each job seeker includes education, work experience, type of job, type of industry, and list of expecting cities. This study considers job seekers who have obtained bachelor's, master's, or doctoral degree. Since most job seekers updated their resume in 2015, here we focus on this majority and include in total 3,443,560 individuals in our analysis.

\begin{figure*}[!htbp]
  \centering
  \includegraphics[width=0.9\textwidth]{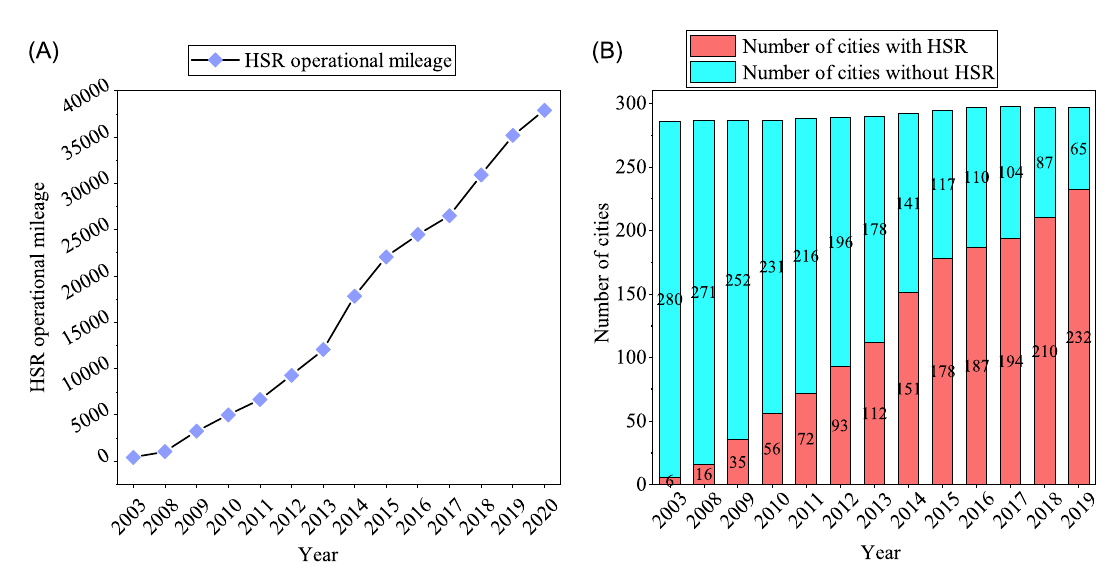}
  \caption{The development of China's HSR network. (A) Increase of HSR operational mileage from 2003 to 2020. (B) Number of cities with HSR (red) and without HSR station (cyan) from 2003 to 2019.}\label{f:station}
\end{figure*}

    \textbf{HSR Data.} We collected opening times, designed speeds, operational speeds, lengths, and stations of Chinese HSR lines. The opening times from 2003 to 2015 are download from the National Railway Administration of the People’s Republic of China (\url{http://www.nra.gov.cn}). Other information is obtained from the Ministry of Transport of the People's Republic of China. Figure \ref{f:station} illustrates the development of the HSR network from 2003 to 2020. By the end of 2020, the operational milage reaches 3,7900 km. The number of cities with at least one HSR station increases from 6 in 2003 to 232 in 2019.

    \textbf{Socioeconomic data.} The socioeconomic factors, such as gross regional product (GRP) and population are mainly obtained from the China City Statistical Yearbook, 2015 (statistical indexes of cities in 2014). Missing data are completed according to the Provincial Statistical Yearbooks and China Regional Statistical Yearbooks.
\subsection{Variables}
    For each job seeker, we approximately estimate the direction of flow which is from living city to expecting city and further aggregate talent flow for city pairs. We then construct a directed and weighted network $G(V, E, W)$, where $V$ is the set of nodes, $E$ is the set of links, and the weight of the link $w_{ij}$ is the volume of talents moving from city $i$ to city $j$.

    We use four measures to quantify flowing patterns, namely, inflow, retention, outflow, and net inflow. The talent inflow ($ f_i^{\text{in}} = \sum_{j \neq i} w_{ji}$) is measured by the number of talents expect to go to city $i$, but live in a different city $j$. The talent retention ($ f_i^{\text{re}} = w_{ii}$) is measured by the number of talents who expect to work in city $i$ where they live. The talent outflow ($f_i^{\mathrm{out}} = \sum_{j \neq i} w_{ij}$) is measured by the number of talents who expect to leave the city $i$ where they live and go to a different city $j$. The net inflow is calculated by the difference between $f_i^{\text{in}}$ and $f_i^{\text{out}}$ in their logarithm forms, say
   $$
        f_i^{\text{net } } = \ln (f_i^{\text{in}}) - \ln ( f_i^{\text{out }}),
   $$
which can be used to evaluate the relative intensity of inflow to outflow. We set HSR as a dummy variable, where a city $i$ served by HSR in 2014  is called a HSR city with HSR$_i=1$, otherwise city $i$ is a non-HSR city with HSR$_i = 0$.

    Following literature on determinants of population migration \citep{shen2012changing, li2014analysis, liu2014spatial,wang2019migration}, we choose some socioeconomic variables related to talent flow as control variables. These variables can be categorized into economic status, population size, investment level, industrial structure, income level, and employment opportunity. (1) Economic status is measured by the gross domestic product per capita (GDPpc, RMB), which is known as an essential factor affecting population migration \citep{lee1966theory,lowry1966migration,wang2020macroeconomic}. In general, the higher the GDPpc is, the stronger attractive the city is. (2) Population size is measured by the total population at the end of the year (Pop, 1 million persons), which is used to reflect the scale of the labor market \citep{liu2014spatial}, and usually positively correlated with population flow. (3) Investment level is measured by the fixed asset investment per capita (FAIpc, RMB). In China, the government plays a critical role in public construction and enterprise development \citep{chen2011government,wang2020exploring}, and thus its investment should be related to talent flow. (4) Industrial structure reflects the level of industrialization and demand for different types of talents \citep{wang2019migration}. Secondary industry plays a vital role in economic development. Hence, industrial development is measured by the GDP of secondary industry per capita (GRPSIpc, RMB). (5) Income level is measured by the wage per capita (Wagepc, RMB). Income is one of major reason behind labor mobility \citep{ravenstein1889law,sjaastad1962costs,wang2019migration}, and it is expected that cities with higher salaries have a stronger capability to retain and attract talents. (6) Employment opportunity is measured by the urban unemployment rate (Uem, \%), where larger unemployment rate indicates less potential need for labors \citep{fang2003migration,cao2018exploring}.

\begin{table}[!htbp]
 \caption{Description of control variables}\label{t:variable}
  \centering
\small
  \begin{tabular*}{\textwidth}{p{0.2\textwidth}p{0.1\textwidth}p{0.6\textwidth}}
   \toprule
  Indicator   & Notation & Description (units) \\
  \midrule
  Economic level & GDPpc & Per capita gross domestic product (RMB) \\
Population &  Pop & Total population at year end (1 million persons)\\
   Investment  & FAIpc & Per capita fixed asset investment (RMB / person), total fixed asset investment / total population at year end \\
  Industrialization level & GRPSIpc & Per capita gross region product of secondary industry (RMB / person), secondary industry value / total population at year end \\
Income level & Wagepc & Per capita wage of staff and workers (RMB) \\
 Job opportunity & Uem & Unemployment rates of urban labour forces (\%), registered unemployed persons in urban areas / persons employed at year end \\
\bottomrule
    \end{tabular*}
\end{table}

\begin{table}[!htbp]
  \centering
   \small
  \caption{Summary statistics.}\label{t:citys}
    \begin{tabular}{p{0.09\textwidth}p{0.065\textwidth}p{0.065\textwidth}p{0.065\textwidth}p{0.065\textwidth}p{0.065\textwidth}p{0.065\textwidth}p{0.065\textwidth}p{0.065\textwidth}p{0.065\textwidth}l@{}}
        \toprule
        Variable	&	Mean	&	&		&	\multicolumn{3}{l}{Standard deviation}	&	Minimum	&	& Maximum & 	\\
        \cmidrule(r){2-4}  \cmidrule(r){5-7}  \cmidrule(r){8-9} \cmidrule(r){10-11}
        	&	non-HSR	&	HSR 	&	All sample		&	non-HSR	&	HSR	&	All sample	&	non-HSR	&	HSR	&	non-HSR	&	HSR	\\
        \midrule
Inflow	&	5.636	&	7.175	&	6.458	&	1.031	&	1.855	&	1.708	&	3.091	&	8.793	&	4.127	&	12.065 \\
Retention	&	5.616	&	7.297	&	6.514	&	1.243	&	2.021	&	1.897	&	3.219	&	10.102	&	3.784	&	12.680 \\
Outflow	&	6.844	&	8.020	&	7.472	&	0.903	&	1.310	&	1.280	&	4.673	&	9.463	&	5.288	&	11.384 \\
Net inflow	&	-1.208	&	-0.845	&	-1.014	&	0.571	&	0.749	&	0.695	&	-2.537	&	0.601	&	-2.274	&	0.889 \\
GDPpc	&	10.513	&	10.796	&	 10.664	&	0.550	&	0.499	&	0.541	&	9.671	&	9.227	&	11.915	&	12.207 \\
Pop	&	1.094	&	1.444	&	1.281	&	0.686	&	0.667	&	0.697	&	-0.942	&	2.515	&	-1.423	&	3.519 \\
FAIpc	&	10.243	&	10.545	&	10.404	&	0.630	&	0.544	&	0.604	&	8.943	&	9.218	&	12.299	&	11.647 \\
GRPSIpc	&	7.749	&	7.792	&	7.772	&	0.856	&	0.875	&	0.865	&	5.270	&	9.940	&	5.411	&	9.671 \\
Wagepc	&	10.741	&	10.800	&	10.772	&	0.168	&	0.203	&	0.189	&	10.211	&	11.366	&	10.317	&	11.546 \\
Uem	&	1.583	&	1.394	&	1.482	&	0.572	&	0.623	&	0.606	&	-0.482	&	3.112	&	-0.960	&	3.269 \\

        \bottomrule
        \multicolumn{11}{@{}p{\textwidth}@{}}{\footnotesize Note: To detect whether there exists multicollinearity, variance inflation factor (VIF) is applied. The VIF of HSR, GDPpc, Pop, FAIpc, GRPSIpc, Wagepc, and Uem is 1.21, 4.99, 1.26, 4.25, 1.10, 1.99, and 1.14, respectively, which are less than the experience values of 10. It suggests the presence of week correlation between explanatory variables or the absence of multicollinearity.}
    \end{tabular}
\end{table}

    Table \ref{t:variable} summarizes the definitions of  control variables. Table \ref{t:citys} shows the descriptive statistics of variables for non-HSR cities, HSR cities, and all cities. All variables are in logarithmic form. On average, compared with non-HSR cities, HSR cities enjoy the higher levels of inflow, retention, outflow, net inflow, Pop, GDPpc, FAIpc, GRPSIpc, and Wagepc, while  have lower Uem.

\subsection{Empirical strategy}

In order to quantify the impact of HSR on the talent flow at city level, we employ a multiple linear regression model which is given by:
\begin{equation}\label{e:q1}
  M_{i} = \beta_0 + \beta_1 \text{HSR}_i + \beta_2 X_{i} + \varepsilon_{i},
\end{equation}
where $M_i$ is the talent flow of city $i$, $X_i$ is the vector of control variables, and $\varepsilon_{i}$ is the error term. $\beta_1$ is the coefficient of interest, which shows the impacts of HSR on talent flow.

There may exist endogeneity problems as many determinants related to talent flow cannot be observed or measured. First, if the setting of regression could not capture related factors effectively, the estimation from the OLS is biased. Second, the governments may have preferences, such as economic level, political value and geographic location in planning HSR lines. Therefore, there may exist selection bias in deciding whether a city is connected to the HSR network, known as selection bias. Third, the good economic status of a city simultaneously increases the attractiveness for talents and the opportunity to be included in the HSR network, resulting in the difficulty in inferring causal relationships between talent flow and the HSR variable.

To identify the causal relationship between transportation infrastructure and socioeconomic development, researchers have adopted a variety of methods such as difference-in-difference (DID) \citep{shao2017high}, propensity score matching (PSM) \citep{jia2017no,long2018high}, and instrumental variable (IV) \citep{dong2018highspeed,zhang2018high,yang2019does,gao2019does,yu2019highspeed}. Among these methods, IV has been widely used to eliminate endogenous concerns when studying the effects of HSR on social and economic development. In particular, one of the most common strategies is simulating the actual layout based on geographical information. Because the hypothetical layout is affected by construction costs and is not only determined by preferences in economy and polities. Following \citet{yang2019does} and \citet{yu2019highspeed}, here we choose the least cost HSR network as IV, where the HSR layout is simulated by the minimum cost spanning tree method \citep{faber2014trade}.

By instrumenting HSR using the least construction cost layout, we use a two-stage least square (2SLS) model to test the robustness of the impact of HSR on talent flow. The 2SLS model is given by:
\begin{equation}\label{eq:iv2}
  \text{HSR}_i = \alpha_{0}+\alpha_{1} \text{PHSR}_{i}+\alpha_{2} X_{i}+\gamma_{i},
\end{equation}
\begin{equation}\label{eq:iv1}
  M_i = \beta_{0} +  \beta_{1} \text{HSR}_{i}+ \beta_2 X_{i}+ \varepsilon_{i},
\end{equation}
where the PHSR is potential HSR connection variable, and $\beta_1$ is the coefficient of interest.
\section{Results}\label{s3}
    To understand how HSR affects talent flow, our empirical framework is composed of four aspects. First, we describe the flowing patterns including inflow, retention, outflow, and net inflow. Second, we use IV constructed by the least cost HSR network to address endogeneity concerns and to test the robustness of our observations. Third, we unpack heterogeneous effects in two aspects: the moderating effects of economic development and the different impacts on different types of industry. Fourth, we check the robustness of major results.
\subsection{Positive impacts of HSR}
\begin{figure*}[htbp]
  \centering
  \includegraphics[width=\textwidth]{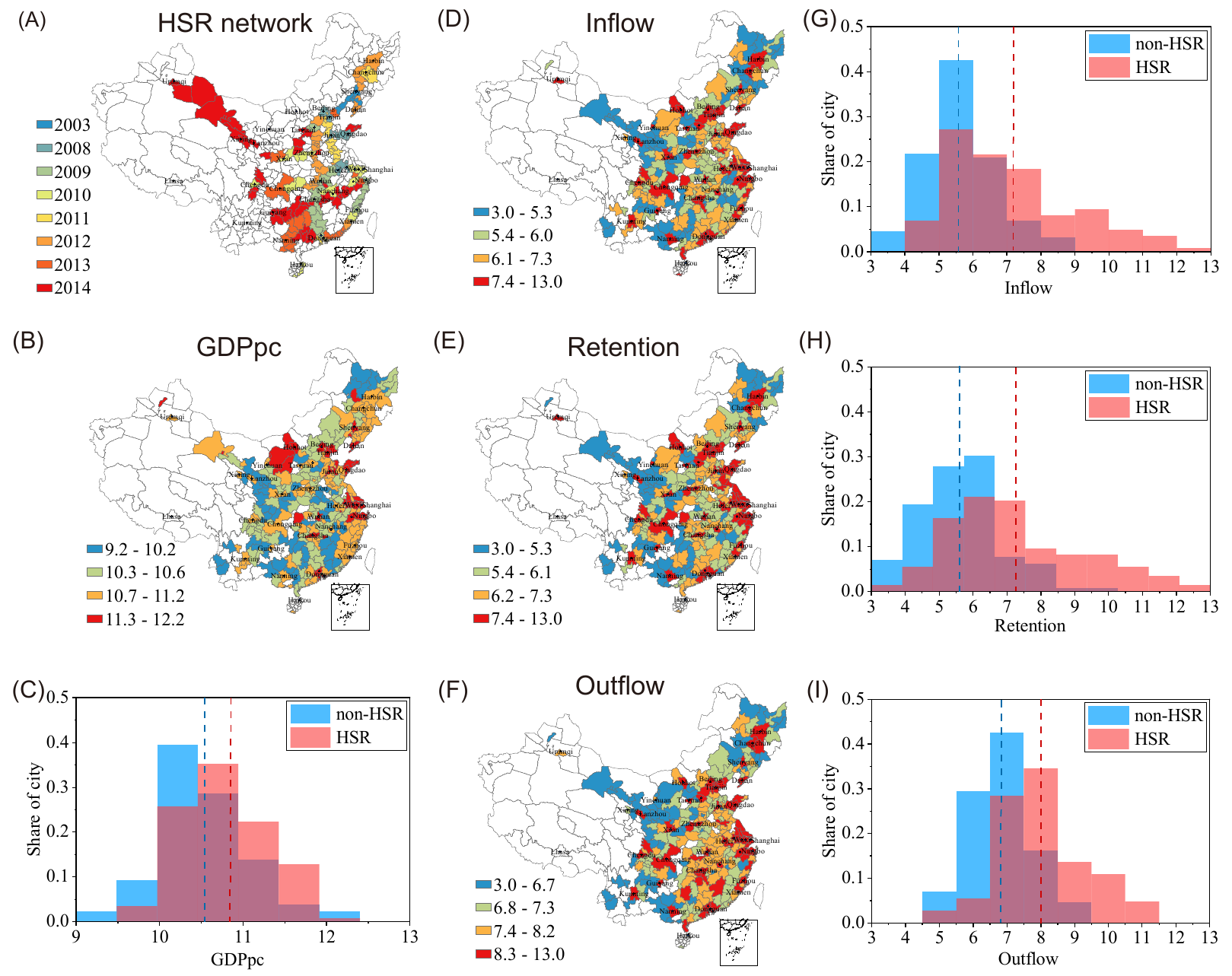}
  \caption{Expansion of HSR network, economic development, and talent flows at the city level. (A) Expansion of HSR from 2003 to 2014. (B) GDPpc of considered cities. (C) Histograms of GDPpc for non-HSR and HSR cities, respectively. (D) - (F) Talent inflows, retentions and outflows of considered cities. (G) - (I) Histograms of inflow, retention and outflow for non-HSR and HSR cities, respectively. The talent inflow, talent retention, talent outflow, and GDPpc are in logarithmic form. Vertical dashed lines mark the means.}\label{t:map}
\end{figure*}

\begin{table}[htbp]
  \centering
  \small
  \caption{Impacts on talent inflow, retention, and outflow.}\label{t:talentflow}
  {
\def\sym#1{\ifmmode^{#1}\else\(^{#1}\)\fi}
\begin{tabular}{p{0.13\textwidth}p{0.12\textwidth}p{0.12\textwidth}p{0.12\textwidth}p{0.13\textwidth}p{0.13\textwidth}l}
\toprule
            &\multicolumn{1}{l}{(1)}&\multicolumn{1}{l}{(2)}&\multicolumn{1}{l}{(3)}&\multicolumn{1}{l}{(4)}&\multicolumn{1}{l}{(5)}&\multicolumn{1}{l}{(6)}\\
\cmidrule(r){2-3}  \cmidrule(r){4-5}  \cmidrule(r){6-7}
&\multicolumn{1}{l}{Inflow}&\multicolumn{1}{l}{Inflow}&\multicolumn{1}{l}{Retention}&\multicolumn{1}{l}{Retention}&\multicolumn{1}{l}{Outflow}&\multicolumn{1}{l}{Outflow}\\
\midrule
HSR         &                     &       0.539\sym{***}&                     &       0.485\sym{***}&                     &       0.399\sym{***}\\
            &                     &     (0.103)         &                     &     (0.107)         &                     &     (0.078)         \\
GDPpc      &       1.374\sym{***}&       1.281\sym{***}&       1.614\sym{***}&       1.530\sym{***}&       1.087\sym{***}&       1.019\sym{***}\\
            &     (0.222)         &     (0.210)         &     (0.224)         &     (0.213)         &     (0.168)         &     (0.160)         \\
Pop       &       1.293\sym{***}&       1.167\sym{***}&       1.635\sym{***}&       1.522\sym{***}&       1.254\sym{***}&       1.161\sym{***}\\
            &     (0.089)         &     (0.088)         &     (0.096)         &     (0.094)         &     (0.068)         &     (0.068)         \\
FAIpc       &       0.298         &       0.199         &       0.426\sym{**} &       0.336\sym{*}  &       0.148         &       0.075         \\
            &     (0.192)         &     (0.172)         &     (0.201)         &     (0.185)         &     (0.145)         &     (0.134)         \\
GRPSIpc      &      -0.164\sym{**} &      -0.162\sym{**} &      -0.211\sym{***}&      -0.209\sym{***}&      -0.080         &      -0.078         \\
            &     (0.070)         &     (0.066)         &     (0.070)         &     (0.067)         &     (0.053)         &     (0.050)         \\
Wagepc     &       2.172\sym{***}&       2.376\sym{***}&       1.863\sym{***}&       2.047\sym{***}&       0.781\sym{***}&       0.932\sym{***}\\
            &     (0.357)         &     (0.355)         &     (0.369)         &     (0.365)         &     (0.285)         &     (0.276)         \\
Uem   &      -0.220\sym{*}  &      -0.189         &      -0.117         &      -0.090         &      -0.065         &      -0.043         \\
            &     (0.127)         &     (0.119)         &     (0.126)         &     (0.119)         &     (0.099)         &     (0.093)         \\
Constant        &     -34.746\sym{***}&     -35.105\sym{***}&     -35.478\sym{***}&     -35.801\sym{***}&     -14.966\sym{***}&     -15.232\sym{***}\\
            &     (3.421)         &     (3.341)         &     (3.414)         &     (3.333)         &     (2.634)         &     (2.550)         \\
\hline
Observation       &         277         &         277         &         277         &         277         &         277         &         277         \\
\textit{F}           &     137.102         &     132.378         &     177.521         &     167.241         &     139.762         &     130.063         \\
RMSE        &       0.827         &       0.790         &       0.824         &       0.795         &       0.633         &       0.607         \\
Adjusted \textit{R$^2$}        &       0.766         &       0.786         &       0.811         &       0.825         &       0.755         &       0.775         \\
\bottomrule
\multicolumn{7}{@{}p{\textwidth}@{}}{ \footnotesize Notes: Variables are in logarithmal form, except for HSR. The coefficients are estimated subject to the ordinary least square (OLS). Robust standard errors are in parentheses. \sym{***},\sym{**} and \sym{*} denote statistical significance at 1\%, 5\%, and 10\%, respectively.}
\end{tabular}
}
\end{table}

     Figure \ref{t:map} (A) illustrates the spatial locations of HSR cities colored by the years they first joined the HSR network, showing an expansion from eastern to western areas. Figure \ref{t:map} (B) shows the distribution of GDPpc in 2014. Figure \ref{t:map} (C) illustrates the histograms of GDPpc of non-HSR and HSR cities. On average, HSR cities enjoy a better economy than non-HSR cities. Figure \ref{t:map} (D) - (F) show the spatial distributions of talent inflows, retentions, and outflows, respectively. Figure \ref{t:map} (G) - (I) illustrate the histograms of inflows, retentions, and outflows of non-HSR and HSR cities, respectively. HSR cities are more likely to attract talents and retain talents than non-HSR cities, while talent outflows of HSR cities are also higher than non-HSR cities.

  Table \ref{t:talentflow} reports the baseline regression for the impact of HSR on talent flows. Columns (1) - (2) present the impacts on talent inflow. HSR increases talent inflow by 53.9\% on average after controlling for socioeconomic variables. At the same time, HSR increases talent retention and outflow by 48.5\%, and 39.9\%, respectively. The regression results further indicate that cities with better economic coditions (e.g., higher GDPpc, higher Wagepc.) are more attractive for talents.
\begin{table}[htbp]
      \centering
    \small
      \caption{Impacts on talent net inflow.}\label{t:netinflow}
      {
    \def\sym#1{\ifmmode^{#1}\else\(^{#1}\)\fi}
    \begin{tabular*}{\linewidth}{p{0.17\linewidth}p{0.17\linewidth}p{0.17\linewidth}p{0.17\linewidth}l@{}}
    \toprule
                &\multicolumn{1}{l}{(1)}&\multicolumn{1}{l}{(2)}&\multicolumn{1}{l}{(3)}&\multicolumn{1}{l}{(4)}\\
   \midrule
    HSR         &       0.362\sym{***}&                     &                     &       0.140\sym{**} \\
                &     (0.079)         &                     &                     &     (0.064)         \\
    GDPpc      &                     &       0.773\sym{***}&       0.287\sym{**} &       0.263\sym{*}  \\
                &                     &     (0.068)         &     (0.142)         &     (0.141)         \\
    Pop       &                     &                     &       0.039         &       0.006         \\
                &                     &                     &     (0.052)         &     (0.052)         \\
    FAIpc       &                     &                     &       0.150         &       0.124         \\
                &                     &                     &     (0.113)         &     (0.109)         \\
    GRPSIpc      &                     &                     &      -0.085\sym{**} &      -0.084\sym{**} \\
                &                     &                     &     (0.037)         &     (0.036)         \\
    Wagepc     &                     &                     &       1.391\sym{***}&       1.444\sym{***}\\
                &                     &                     &     (0.211)         &     (0.216)         \\
    Uem   &                     &                     &      -0.154\sym{***}&      -0.147\sym{***}\\
                &                     &                     &     (0.051)         &     (0.050)         \\
    Constant      &      -1.208\sym{***}&      -1.014\sym{***}&     -19.781\sym{***}&     -19.874\sym{***}\\
            &     (0.050)         &     (0.033)         &     (1.926)         &     (1.934)         \\
    \midrule
   Observation       &         277         &         277         &         277         &         277         \\
    \textit{F}           &      20.783         &     130.078         &      50.598         &      45.222         \\
    RMSE        &       0.672         &       0.556         &       0.496         &       0.493         \\
    Adjusted \textit{R$^2$  }      &       0.065         &       0.361         &       0.491         &       0.497         \\
    \bottomrule
    \multicolumn{5}{@{}p{\linewidth}@{}}{\footnotesize Notes: Variables are in logarithmal form, except for HSR. Robust standard errors are in parentheses. \sym{***}, \sym{**}, and \sym{*} denote statistical significance at 1\%, 5\%, and 10\%, respectively.}\\

\end{tabular*}
}
\end{table}

\begin{figure*}[htbp]
  \centering
  \includegraphics[width=\linewidth]{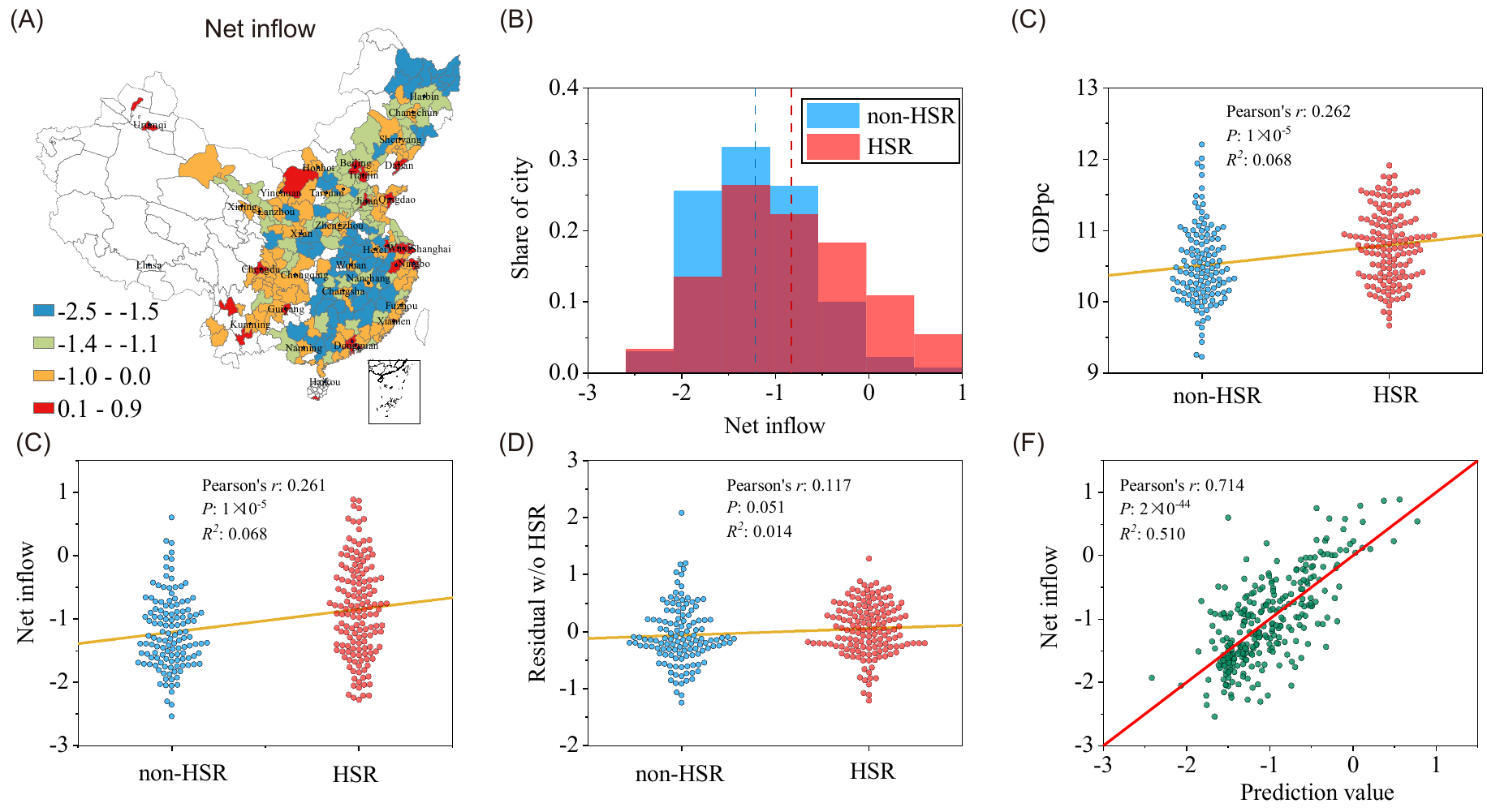}
  \caption{Pattern of net inflow. (A) Distribution of net inflow at the city level. (B) Histogram of net inflows of non-HSR and HSR cities. (C) GDPpc of non-HSR and HSR cities. (D) Net inflows of non-HSR and HSR cities. (E) Residuals of regression without HSR of non-HSR and HSR cities. (F) Relationship between predicting and real net inflows. Yellow and red line represents regression line. }\label{f:netinflow}
\end{figure*}

    Figure \ref{f:netinflow} (A) illustrates the spatial distribution of net inflows, showing that most cities are losing talents. Figure \ref{f:netinflow} (B) shows the histograms of net inflows for non-HSR and HSR cities where we find that HSR cities have a lower level of talent loss than non-HSR cities (see also column (1) of Table \ref{t:netinflow}). As shown in Fig. \ref{f:netinflow} (C), most HSR cities is richer than most non-HSR cities (Pearson's \textit{r} = 0.262, \textit{P} < 0.01). Similar to distributions of GDPpc of non-HSR and HSR cities, as shown in Fig. \ref{f:netinflow} (D), the average net inflow of non-HSR cities is lower than that of HSR cities (Pearson's \textit{r} = 0.261, \textit{P} < 0.01). Column (2) of Table \ref{t:netinflow} reports the result only controlling for GDPpc. The coefficient of GDPpc is positive with statistical significance at 1\% level, indicating that a city with high economic status attracts more talents. After adding more socioeconomic factors, the regression shows that GDPpc, Pop, and Wagepc are positively correlated with talent net inflow. Figure \ref{f:netinflow} (E) shows the positive correlation between HSR and residuals of the regression without HSR. suggesting a significant explanatory power. Column (4) reports the regression result with controlling for both HSR and socioeconomic factors. When a city is connected to the HSR network, the inflow-outflow ratio will increase by 14\% on average. As shown in Fig. \ref{t:netinflow} (F), all variables can explain up to 51\% of the variance, with the correlation between actual and predicting values being 0.714.

\subsection{Analysis using an instrumental variable}
\begin{table}[htbp]
      \centering
      \small
    \caption{Results from IV method.}\label{t:baselineiv}
    {
    \def\sym#1{\ifmmode^{#1}\else\(^{#1}\)\fi}
    \begin{tabular}{p{0.15\textwidth}p{0.14\textwidth}p{0.14\textwidth}p{0.14\textwidth}p{0.14\textwidth}p{0.14\textwidth}l@{}}
    \toprule
                &\multicolumn{1}{l}{(1)}&\multicolumn{1}{l}{(2)}&\multicolumn{1}{l}{(3)}&\multicolumn{1}{l}{(4)}&\multicolumn{1}{l}{(5)}\\
    \cmidrule(r){2-2}     \cmidrule(r){3-6}
     &\multicolumn{1}{l}{HSR}&\multicolumn{1}{l}{Inflow}&\multicolumn{1}{l}{Retention}&\multicolumn{1}{l}{Outflow}&\multicolumn{1}{l}{Net inflow}\\
    \midrule
    PHSR        &       0.330\sym{***}&                     &                     &                     &                     \\
                &     (0.061)         &                     &                     &                     &                     \\
    HSR         &                     &       1.468\sym{***}&       1.424\sym{***}&       0.988\sym{***}&       0.480\sym{**} \\
                &                     &     (0.372)         &     (0.367)         &     (0.278)         &     (0.209)         \\
    GDPpc      &       0.239\sym{**} &       1.122\sym{***}&       1.369\sym{***}&       0.917\sym{***}&       0.204         \\
                &     (0.099)         &     (0.224)         &     (0.227)         &     (0.173)         &     (0.140)         \\
    Constant       &       2.426         &     -35.725\sym{***}&     -36.427\sym{***}&     -15.624\sym{***}&     -20.101\sym{***}\\
            &     (1.543)         &     (3.667)         &     (3.643)         &     (2.682)         &     (2.022)         \\
    \midrule
    Observation       &         277         &         277         &         277         &         277         &         277         \\
    \textit{F}           &      24.622         &     119.127         &     149.973         &     119.745         &      43.190         \\
    RMSE        &       0.435         &       0.885         &       0.891         &       0.655         &       0.509         \\
    Adjusted \textit{R$^2$}        &       0.241         &       0.723         &       0.773         &       0.730         &       0.447         \\
    \bottomrule
    \multicolumn{6}{@{}p{\textwidth}@{}@{}}{\footnotesize Notes: Variables are in logarithmal form, except for HSR. PHSR is potential HSR connection variable. Robust standard errors are in parentheses. \sym{***}, \sym{**}, and \sym{*} denote statistical significance at 1\%, 5\%, and 10\%, respectively.}
    \end{tabular}
    }
\end{table}

    In order to address the endogeneity of HSR lines, we use the least cost HSR network as IV. Table \ref{t:baselineiv} presents the 2SLS regression results. Column (1) presents the first stage estimation result. The value of  Kleibergen-Paap rk \textit{LM} statistic is 25.520 and significant at 1\% level, suggesting that IV is identified. The value of Kleibergen-Paap rk Wald \textit{F} statistic is 29.525, greater than the experience value of 10, indicating the pass of weak identification test. Columns (2) - (4) present the 2SLS estimation results. The coefficients of HSR are 1.468, 1.424, and 0.988, all of which are significant at 1\% level. The results indicate that to join the HSR network will increase the ability to retain talents and promote the talent flow, which is consistent with the baseline regression results. Column (5) reports the impacts on net inflow. The coefficient of HSR is 0.480 and significant at 5\% level, indicating that HSR has positive impacts on net inflow after eliminating endogeneity. The coefficients from the 2SLS are larger than those of OLS estimation, indicating that the endogeneity leads to an underestimation of the impact of HSR on talent flow.

\subsection{Heterogeneous effects of HSR}
\begin{table}[htbp]
      \centering
\small
      \caption{Moderating effects of economic development.}\label{t:moderation}
      {
    \def\sym#1{\ifmmode^{#1}\else\(^{#1}\)\fi}
    \begin{tabular*}{\textwidth}{p{0.2\linewidth}p{0.17\linewidth}p{0.17\linewidth}p{0.17\linewidth}l@{}}
    \toprule
                &\multicolumn{1}{l}{(1)}&\multicolumn{1}{l}{(2)}&\multicolumn{1}{l}{(3)}&\multicolumn{1}{l}{(4)}\\
            \cmidrule(r){2-5}
 &\multicolumn{1}{l}{Inflow}&\multicolumn{1}{l}{Retention}&\multicolumn{1}{l}{Outflow}&\multicolumn{1}{l}{Net inflow}\\
    \midrule
    HSR         &       0.571\sym{***}&       0.515\sym{***}&       0.416\sym{***}&       0.155\sym{**} \\
            &     (0.098)         &     (0.103)         &     (0.078)         &     (0.061)         \\
    GDPpc      &       0.811\sym{***}&       1.086\sym{***}&       0.773\sym{***}&       0.038         \\
                &     (0.188)         &     (0.205)         &     (0.156)         &     (0.143)         \\
   HSR$\times$GDPpc&       1.240\sym{***}&       1.171\sym{***}&       0.647\sym{***}&       0.593\sym{***}\\
                &     (0.175)         &     (0.193)         &     (0.152)         &     (0.110)         \\
     Pop       &       1.123\sym{***}&       1.481\sym{***}&       1.138\sym{***}&      -0.015         \\
                &     (0.078)         &     (0.082)         &     (0.062)         &     (0.052)         \\
    FAIpc       &       0.145         &       0.286\sym{*}  &       0.047         &       0.098         \\
                &     (0.149)         &     (0.158)         &     (0.120)         &     (0.110)         \\
    GRPSIpc      &      -0.154\sym{***}&      -0.202\sym{***}&      -0.074         &      -0.081\sym{**} \\
                &     (0.057)         &     (0.058)         &     (0.046)         &     (0.035)         \\
    Wagepc     &       2.050\sym{***}&       1.739\sym{***}&       0.762\sym{***}&       1.288\sym{***}\\
                &     (0.320)         &     (0.333)         &     (0.272)         &     (0.201)         \\
    Uem   &      -0.125         &      -0.029         &      -0.009         &      -0.116\sym{**} \\
                &     (0.103)         &     (0.105)         &     (0.085)         &     (0.046)         \\
     Constant      &       6.066\sym{***}&       6.156\sym{***}&       7.205\sym{***}&      -1.139\sym{***}\\
            &     (0.063)         &     (0.077)         &     (0.057)         &     (0.042)         \\
    \midrule
    Observation       &         277         &         277         &         277         &         277         \\
    \textit{F}           &     187.162         &     221.118         &     146.664         &      61.991         \\
    RMSE        &       0.725         &       0.737         &       0.585         &       0.469         \\
    Adjusted \textit{R$^2$}        &       0.820         &       0.849         &       0.791         &       0.544         \\
    \bottomrule
    \multicolumn{5}{@{}p{\linewidth}}{\footnotesize Notes: Variables are in logarithmal form and mean-centered, except for HSR. Robust standard errors are in parentheses. \sym{***}, \sym{**}, and \sym{*} denote statistical significance at 1\%, 5\%, and 10\%, respectively.} \\
    \end{tabular*}
    }
\end{table}
\begin{figure}[htbp]
  \centering
  \includegraphics[width=0.75\linewidth]{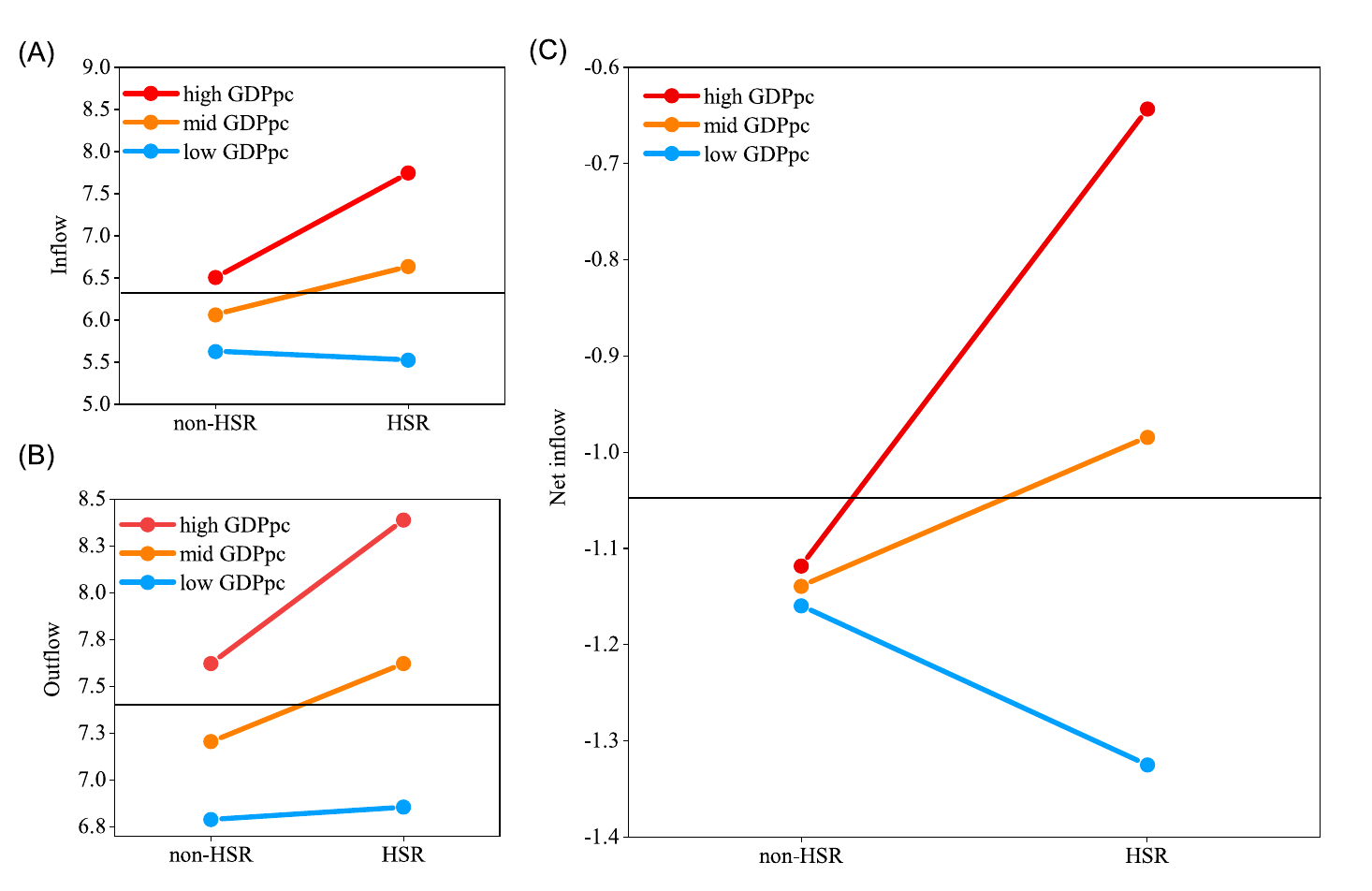}
  \caption{Moderating effects of economic status on the relationship between HSR and talent flow. (A) Talent inflow. (B) Talent outflow. (C) Talent net inflow. Horizontal line represents regression constant. Low and high GDPpc means cities on average have GDPpc that is 1 SD below and above the mean value 0 (mid GDPpc), respectively.}\label{f:moderation}
\end{figure}

We first discuss the different effects of HSR on cities with different economic development levels \citep{li2020economic}. To do so, we introduce an interaction term between HSR and GDPpc, say HSR $\times$ GDPpc into the regression models (see Table \ref{t:moderation}). Both coefficients of HSR and HSR $\times$ GDPpc are positive and significant at 1\% level, indicating that HSR strengthens inflow, retention, and outflow in economically developed cities. In other words, HSR contributes more to cities with better economic status.

    Based on the regression estimations of Table \ref{t:moderation}, the moderating effects are visualized in Fig. \ref{f:moderation}. HSR in cities with higher GDPpc promotes both inflow and outflow, while for cities with low GDPpc, HSR cities have even lower inflow. This effect becomes more pronounced when considering the net inflow. Firstly, average net inflows of non-HSR cities with different GDPpc level are more or less the same, whereas there are significant gaps between HSR cities in different economic status. Secondly, HSR in cities with high GDPpc (1 SD above the mean) are much more attractive for talents than non-HSR cities, while HSR cities with low GDPpc (1 SD below the mean) lose talents more quickly than non-HSR cities. That is to say, HSR marginally increases net inflow in developed cities but leads to a larger brain drain in less-developed cities. Such phenomenon, known as the siphoning effects of HSR, is consistent with previous results in \citet{qin2017no}, \citet{ke2017china}, \citet{yu2019highspeed}, and \citet{gao2020does}. These results provide supportive evidence for some arguments in the filed of the new economic geography that the reduction of transport costs will strengthen the core-periphery structure and lead to economic polarization \citep{helpman1985market,krugman1991increasing}.

    \begin{table}[!htbp]
      \caption{ Heterogeneous analysis from IV method.}\label{t:industyIV}
      \centering
      \small
    {
    \def\sym#1{\ifmmode^{#1}\else\(^{#1}\)\fi}
    \begin{tabular}{p{0.13\textwidth}p{0.10\textwidth}p{0.10\textwidth}p{0.10\textwidth}p{0.14\textwidth}p{0.15\textwidth}p{0.14\textwidth}l}
    \toprule
                           &(1) & (2) & (3) & (4) & (5)& (6)\\
    \cmidrule(r){2-4}\cmidrule(r){5-7}
                              &Primary&Secondary   & Tertiary  & Producer service  & Consumer service &Public service \\
    \midrule
    \multicolumn{3}{l}{Panel A : Inflow}\\
    HSR          &       1.429\sym{***}&       1.546\sym{***}&       1.376\sym{***}&       1.389\sym{***}&       1.388\sym{***}&       1.269\sym{***}\\
                &     (0.383)         &     (0.383)         &     (0.367)         &     (0.374)         &     (0.378)         &     (0.350)         \\

    Adjusted \textit{R$^{2}$ }       &       0.521         &       0.713         &       0.719         &       0.723         &       0.697         &       0.693         \\
    \midrule
     \multicolumn{3}{l}{Panel B: Retention}\\
    HSR           &       1.303\sym{***}&       1.511\sym{***}&       1.365\sym{***}&       1.394\sym{***}&       1.312\sym{***}&       1.237\sym{***}\\
                &     (0.365)         &     (0.381)         &     (0.359)         &     (0.362)         &     (0.377)         &     (0.350)         \\
    Adjusted \textit{R$^{2}$}       &       0.540         &       0.767         &       0.766         &       0.764         &       0.764         &       0.738         \\

    \midrule

    \multicolumn{3}{l}{Panel C: Outflow}\\
    HSR         &       0.943\sym{**} &       1.009\sym{***}&       0.938\sym{***}&       0.982\sym{***}&       0.856\sym{***}&       0.796\sym{***}\\
                &     (0.370)         &     (0.290)         &     (0.277)         &     (0.282)         &     (0.288)         &     (0.294)         \\
    Adjusted \textit{R$^{2}$}       &       0.462         &       0.726         &       0.727         &       0.720         &       0.726         &       0.685         \\
    \midrule

    \multicolumn{3}{l}{Panel D:  Net inflow}\\
    HSR         &       0.486         &       0.537\sym{**} &       0.438\sym{**} &       0.407\sym{*}  &       0.532\sym{**} &       0.473\sym{*}  \\
                &     (0.338)         &     (0.218)         &     (0.209)         &     (0.214)         &     (0.230)         &     (0.256)         \\
    Adjusted\textit{ R$^{2}$}         &       0.140         &       0.421         &       0.419         &       0.442         &       0.319         &       0.264         \\

    \bottomrule
    \multicolumn{7}{@{}p{\textwidth}@{}}{\footnotesize Notes: Variables are in logarithmal form, except for HSR. Robust standard errors are in parentheses. \sym{***}, \sym{**}, and \sym{*} denote statistical significance at 1\%, 5\%, and 10\%, respectively. The producer service industry includes transportation, warehousing, postal, information transmission, computer services and software, finance, insurance, leasing, business services, scientific research, and technical services. The consumer service industry includes wholesale, retail trade, hotels, catering, real estate, culture, sports, and entertainment. The public service industry includes education, health, social work, welfare, public administration, and social organization.}
    \end{tabular}
    }
    \end{table}

Next, we test whether the effects of HSR on different types of talents are different. According to type of industrial sectors which their jobs are in, we categorize talents into primary industry, secondary industry, and tertiary industry groups. In particular, the tertiary industry contains many sectors of different nature. \citet{shao2017high} and \citet{tian2021impact} find that the effects of HSR on the agglomeration of different sectors in the tertiary industry are heterogeneous. The reasons for it are as follows. Producer service is a skill-intensive industry, relying on technology, knowledge, and human capital. As HSR can promote the diffusion of knowledge, information, and experience, HSR may have positive effects on the flow of talents in the producer service subgroup. Consumer service relies on face-to-face communication, obviously benefitting from HSR, and thus HSR may produce positive effects on the flow of talents in the consumer service subgroup. Therefore, we further divide the tertiary industry group into producer service, consumer service, and public service subgroups. Panel A, Panel B, Panel C, and Panel D in Table \ref{t:industyIV} present the heterogeneous impacts of HSR on inflow, retention, outflow, and net inflow, respectively. The effects of HSR on the flow of talents in secondary group, producer service subgroup, and consumer service subgroup are larger. It suggests the role of HSR in the exchange of knowledge, technology, skill, information, and other production factors.

\subsection{Robustness checks}\label{s4}
\begin{table}[!htbp]
  \centering
  \small
    \caption{Robustness checks with different treatment groups estimated from OLS.}\label{t:check}
  {
    \def\sym#1{\ifmmode^{#1}\else\(^{#1}\)\fi}
   \begin{tabular}{p{0.12\textwidth}p{0.08\textwidth}p{0.08\textwidth}p{0.08\textwidth}p{0.1\textwidth}p{0.08\textwidth}p{0.08\textwidth}p{0.08\textwidth}l}
    \toprule
                &\multicolumn{1}{l}{(1)}&\multicolumn{1}{l}{(2)}&\multicolumn{1}{l}{(3)}&\multicolumn{1}{l}{(4)}
&\multicolumn{1}{l}{(5)}&\multicolumn{1}{l}{(6)}&\multicolumn{1}{l}{(7)}&\multicolumn{1}{l}{(8)}\\
\cmidrule(r){2-5} \cmidrule(r){6-9}
     & 2012 & & & & 2013 & & &   \\
\cmidrule(r){2-5} \cmidrule(r){6-9}
                &Inflow  &Retention & Outflow  & Net inflow & Inflow  &Retention & Outflow  & Net inflow\\
    \midrule
    HSR  & 0.479\sym{***} &  0.495\sym{***}&  0.428\sym{***}&  0.051  &0.529\sym{***}  & 0.468\sym{***} & 0.387\sym{***}&  0.142\sym{**} \\
                 &   (0.129)    &     (0.126)    &   (0.094)       & (0.073)  &  (3.849)    &     (3.828)     & (2.984)    &  (2.474)   \\
    Adjusted \textit{R$^{2}$ }      &      0.778             &         0.822    &       0.773         &       0.490         &     0.782      &        0.822       &       0.771         &       0.496         \\
    \bottomrule
    \multicolumn{9}{@{}p{\textwidth}@{}}{\footnotesize Notes: All regression results control for control variables. Robust standard errors are in parentheses. \sym{***}, \sym{**}, and \sym{*} denote statistical significance at 1\%, 5\%, and 10\%, respectively.}
    \end{tabular}
    }
\end{table}

    In the baseline regression model, the treatment group consists of cities that are serviced by HSR in 2014. As there may be a time-lag in obtaining information and making decisions for job seekers, we choose cities with HSR stations in 2012 and 2013 as treatment groups for robustness checks. If the coefficients of HSR are significant and positive, it indicates that HSR has a long-term impact on talent flow and thus supports the primary findings. Columns (1) - (4) and (5) - (8) in Table \ref{t:check} are the results responding to different treatment groups in 2012 and 2013, respectively. The effects of HSR are positive, indicating that HSR has impacts on talent flow being consistent with the baseline results in Table \ref{t:talentflow} and Table \ref{t:netinflow}.

\begin{table}[!htbp]
\small
  \caption{Heterogeneous analysis from IV method using 2012 and 2013 HSR network.}\label{t:industryIV2012}
  \centering
  \small
{
\def\sym#1{\ifmmode^{#1}\else\(^{#1}\)\fi}
\begin{tabular}{p{0.13\textwidth}p{0.12\textwidth}p{0.10\textwidth}p{0.10\textwidth}p{0.15\textwidth}p{0.15\textwidth}l@{}}
\toprule
                       &(1) & (2) & (3) & (4) & (5)& (6)\\
\cmidrule(r){2-7}
                          &Primary &Secondary  & Tertiary  & Producer service  & Consumer service &Public service \\
\midrule
\multicolumn{3}{l}{Panel A: Inflow}\\
HSR (2012)     &       2.868\sym{**} &       3.103\sym{***}&       2.762\sym{**} &       2.788\sym{**} &       2.786\sym{**} &       2.547\sym{**} \\
            &     (1.171)         &     (1.204)         &     (1.127)         &     (1.146)         &     (1.121)         &     (1.100)         \\

Adjusted \textit{R$^{2}$ }        &       0.121         &       0.398         &       0.444         &       0.451         &       0.437         &       0.394         \\
HSR (2013)     &       1.888\sym{***}&       2.043\sym{***}&       1.819\sym{***}&       1.836\sym{***}&       1.834\sym{***}&       1.677\sym{***}\\
            &     (0.612)         &     (0.617)         &     (0.583)         &     (0.595)         &     (0.592)         &     (0.562)         \\

Adjusted \textit{R$^{2}$}        &        0.420         &       0.645         &       0.658         &       0.663         &       0.640         &       0.625         \\
\midrule
 \multicolumn{3}{l}{Panel B: Retention}\\
HSR (2012)     &       2.616\sym{**} &       3.034\sym{***}&       2.739\sym{**} &       2.798\sym{**} &       2.633\sym{**} &       2.484\sym{**} \\
            &     (1.076)         &     (1.176)         &     (1.090)         &     (1.120)         &     (1.062)         &     (1.019)         \\
Adjusted \textit{R$^{2}$}       &      0.221         &       0.533         &       0.563         &       0.549         &       0.602         &       0.530         \\
HSR (2013)     &       1.723\sym{***}&       1.998\sym{***}&       1.804\sym{***}&       1.842\sym{***}&       1.734\sym{***}&       1.636\sym{***}\\
            &     (0.573)         &     (0.607)         &     (0.570)         &     (0.581)         &     (0.573)         &     (0.543)         \\

Adjusted \textit{R$^{2}$}       &      0.465         &       0.711         &       0.718         &       0.711         &       0.730         &       0.688         \\
\midrule

\multicolumn{3}{l}{Panel C: Outflow}\\
HSR (2012)     &       1.893\sym{**} &       2.026\sym{**} &       1.882\sym{**} &       1.971\sym{**} &       1.718\sym{**} &       1.597\sym{**} \\
            &     (0.946)         &     (0.825)         &     (0.771)         &     (0.800)         &     (0.741)         &     (0.760)         \\

Adjusted \textit{R$^{2}$}       &       0.251         &       0.520         &       0.540         &       0.519         &       0.576         &       0.541         \\
HSR (2013)     &       1.246\sym{**} &       1.334\sym{***}&       1.239\sym{***}&       1.298\sym{***}&       1.131\sym{***}&       1.052\sym{**} \\
            &     (0.548)         &     (0.443)         &     (0.421)         &     (0.433)         &     (0.420)         &     (0.435)         \\

Adjusted \textit{R$^{2}$}        &       0.396         &       0.673         &       0.677         &       0.667         &       0.683         &       0.639         \\
\midrule
\multicolumn{3}{l}{Panel D: Net inflow}\\
HSR (2012)     &       0.975         &       1.078\sym{*}  &       0.880\sym{*}  &       0.817         &       1.068\sym{*}  &       0.949         \\
            &     (0.740)         &     (0.555)         &     (0.524)         &     (0.521)         &     (0.583)         &     (0.628)         \\
Adjusted \textit{R$^{2}$}        &     0.020         &       0.156         &       0.200         &       0.259         &       0.066         &       0.021         \\
HSR (2013)     &       0.642         &       0.710\sym{**} &       0.579\sym{*}  &       0.538\sym{*}  &       0.703\sym{**} &       0.625\sym{*}  \\
            &     (0.455)         &     (0.316)         &     (0.299)         &     (0.303)         &     (0.333)         &     (0.361)         \\

Adjusted \textit{R$^{2}$}       &         0.122         &       0.379         &       0.386         &       0.413         &       0.284         &       0.230         \\
\bottomrule
\multicolumn{7}{@{}p{\textwidth}@{}}{\footnotesize Notes: All regression results control for control variables. Robust standard errors are in parentheses. \sym{***}, \sym{**}, and \sym{*} denote statistical significance at 1\%, 5\%, and 10\%, respectively.}\\
\end{tabular}
}
\end{table}
     Table \ref{t:industryIV2012} reports the 2SLS regression results using the HSR networks in 2012 and 2013, respectively. The results indicate that HSR significantly affects the flow of talents in primary, secondary, and tertiary group, supporting the robustness of these results in Table \ref{t:industyIV}.

\section{Conclusions and discussion}\label{s5}

     Well-educated workers play a vital role in regional economic development. As the quality and quantity of the economy have increased, there is a rising demand for workers with high-skilled and professional knowledge, thus intensifying competitiveness for talents among cities. Since HSR improves accessibility and decreases travel time  \citep{fang2003migration,fan2005modeling,shen2012changing,li2014analysis,cao2018exploring}, it is expected that HSR affects the flow of talents. Based on large-scale resume data of Chinese online job seekers, this work conveys three major findings. (1) HSR has significant and positive effects on talent inflow, retention, outflow, and net inflow. It suggests that the decrease in travel time encourages the flow of talents between cities. (2) We find the moderating role of economic development in the effects of HSR on talent flow. Developed cities gain talents via HSR while less-developed cities lose talents because of the siphoning effects \citep{amos2010high,yu2019highspeed,gao2020does}. (3) HSR has greater impacts on secondary, producer service, and consumer service industry-specific talents, supporting that HSR connection has a higher correlation with the development of knowledge-intensive, technology-intensive, and communication-intensive industries \citep{lin2017travel, shao2017high, dong2018highspeed, tian2021impact}.

    Readers should be aware of some limitations of this study. (1) We treat HSR as a dummy variable in the regression model, which ignores some important characteristics of HSR network across cities, such as the number of HSR lines. Also, the impacts of HSR on cities with different levels of transport capacity may be different. (2) The online job seekers only account for a small part of all job seekers, so our resume data may not perfectly describe talent flow in China. (3) With the development of society and economy, the trend of talent flow will evolve over time. Here, we only analyze job seekers who update resumes in 2015, which could not reflect the latest status. In addition, with more and more peripheral cities being connected to the HSR network \citep{jiao2014impacts}, the impacts of HSR on talent flow may change, so it is valuable to find timely and high-resolution data to capture the temporal trends. (4) We do not consider the motivation of talent flow as we do not have sufficient data. There are complex mechanisms behind talent flow, and it is valuable to further explore the impacts of HSR on population flow, combining with personal and household characteristics, such as gender, earning level, occupation, and household structure \citep{huang2018tracking}. Thanks to the development of informational techniques and the emergence of various platforms, a large amount of multi-dimensional data related to human activities has been tracked and recorded \citep{gao2019computational,zhou2021representative}. It thus provides rich opportunities in studying the socioeconomic impacts of HSR.

    Our results have important policy implications for talent recruitment/management and urban planning. First, our results indicate that less-developed cities with HSR will lose talents due to the siphoning effects. In addition, HSR greatly boosts the development of tertiary industry. If a less-developed city wants to benefit from HSR projects, the government should designedly change the industrial structure to match with HSR, thus enhance competitiveness in attracting talents. Second, policymakers should make full use of the advantage of tourism resources, and improve the service quality of catering, accommodation, and transportation. Then, job opportunities and economic environment can be improved through HSR construction. Third, government could encourage communications and cooperations among cities to promote technological innovations and knowledge diffusion, and thus peripheral cities could learn from other cities, especially from core cities.

\section*{Acknowledgements}
We thank Qinyuan Wu and Wei Bai for collecting data, and Zhongtao Yue and Yan-li Lee for cleaning the raw data. This study was partially supported by the National Natural Science Foundation of China (Grant No. 11975071).
\section*{Author Contributions}
    MX, JG, and TZ design the research, analyze the data, and prepare the manuscript. MX perform the research.
\section*{Conflicts of Interest}
The authors declare no conflict of interest.
\small
\bibliographystyle{cas-model2-names}
 \bibliography{cityandtalent}

\begin{thebibliography}{74}
\expandafter\ifx\csname natexlab\endcsname\relax\def\natexlab#1{#1}\fi
\providecommand{\url}[1]{\texttt{#1}}
\providecommand{\href}[2]{#2}
\providecommand{\path}[1]{#1}
\providecommand{\DOIprefix}{doi:}
\providecommand{\ArXivprefix}{arXiv:}
\providecommand{\URLprefix}{URL: }
\providecommand{\Pubmedprefix}{pmid:}
\providecommand{\doi}[1]{\href{http://dx.doi.org/#1}{\path{#1}}}
\providecommand{\Pubmed}[1]{\href{pmid:#1}{\path{#1}}}
\providecommand{\bibinfo}[2]{#2}
\ifx\xfnm\relax \def\xfnm[#1]{\unskip,\space#1}\fi
\bibitem[{Alonso(1964)}]{alonso1964location}
\bibinfo{author}{Alonso, W.}, \bibinfo{year}{1964}.
\newblock \bibinfo{title}{{Location and land use: Toward a general theory of
  land rent}}.
\newblock \bibinfo{publisher}{Cambridge, USA: Harvard University Press}.
\bibitem[{Amos et~al.(2010)Amos, Bullock and Sondhi}]{amos2010high}
\bibinfo{author}{Amos, P.}, \bibinfo{author}{Bullock, D.},
  \bibinfo{author}{Sondhi, J.}, \bibinfo{year}{2010}.
\newblock \bibinfo{title}{{High-speed rail: The fast track to economic
  development?}}
\newblock \bibinfo{publisher}{Beijing, China: World Bank}.
\bibitem[{Campa et~al.(2016)Campa, L{\'o}pez-Lambas and Guirao}]{campa2016high}
\bibinfo{author}{Campa, J.L.}, \bibinfo{author}{L{\'o}pez-Lambas, M.E.},
  \bibinfo{author}{Guirao, B.}, \bibinfo{year}{2016}.
\newblock \bibinfo{title}{{High speed rail effects on tourism: Spanish
  empirical evidence derived from China's modelling experience}}.
\newblock \bibinfo{journal}{Journal of Transport Geography}
  \bibinfo{volume}{57}, \bibinfo{pages}{44--54}.
\bibitem[{Cao et~al.(2018)Cao, Zheng, Liu, Li and Chen}]{cao2018exploring}
\bibinfo{author}{Cao, Z.}, \bibinfo{author}{Zheng, X.}, \bibinfo{author}{Liu,
  Y.}, \bibinfo{author}{Li, Y.}, \bibinfo{author}{Chen, Y.},
  \bibinfo{year}{2018}.
\newblock \bibinfo{title}{{Exploring the changing patterns of China's migration
  and its determinants using census data of 2000 and 2010}}.
\newblock \bibinfo{journal}{Habitat International} \bibinfo{volume}{82},
  \bibinfo{pages}{72--82}.
\bibitem[{Cascetta et~al.(2011)Cascetta, Papola, Pagliara and
  Marzano}]{cascetta2011analysis}
\bibinfo{author}{Cascetta, E.}, \bibinfo{author}{Papola, A.},
  \bibinfo{author}{Pagliara, F.}, \bibinfo{author}{Marzano, V.},
  \bibinfo{year}{2011}.
\newblock \bibinfo{title}{{Analysis of mobility impacts of the high speed
  Rome--Naples rail link using withinday dynamic mode service choice models}}.
\newblock \bibinfo{journal}{Journal of Transport Geography}
  \bibinfo{volume}{19}, \bibinfo{pages}{635--643}.
\bibitem[{Chen and Hall(2011)}]{chen2011impacts}
\bibinfo{author}{Chen, C.L.}, \bibinfo{author}{Hall, P.}, \bibinfo{year}{2011}.
\newblock \bibinfo{title}{{The impacts of high-speed trains on British economic
  geography: A study of the UK's InterCity} 125/225 and its effects}.
\newblock \bibinfo{journal}{Journal of Transport Geography}
  \bibinfo{volume}{19}, \bibinfo{pages}{689--704}.
\bibitem[{Chen and Hall(2012)}]{chen2012wider}
\bibinfo{author}{Chen, C.L.}, \bibinfo{author}{Hall, P.}, \bibinfo{year}{2012}.
\newblock \bibinfo{title}{{The wider spatial-economic impacts of high-speed
  trains: a comparative case study of Manchester and Lille sub-regions}}.
\newblock \bibinfo{journal}{Journal of Transport Geography}
  \bibinfo{volume}{24}, \bibinfo{pages}{89--110}.
\bibitem[{Chen et~al.(2011)Chen, Sun, Tang and Wu}]{chen2011government}
\bibinfo{author}{Chen, S.}, \bibinfo{author}{Sun, Z.}, \bibinfo{author}{Tang,
  S.}, \bibinfo{author}{Wu, D.}, \bibinfo{year}{2011}.
\newblock \bibinfo{title}{{Government intervention and investment efficiency:
  Evidence from China}}.
\newblock \bibinfo{journal}{Journal of Corporate Finance} \bibinfo{volume}{17},
  \bibinfo{pages}{259--271}.
\bibitem[{Cheng et~al.(2015)Cheng, Loo and Vickerman}]{cheng2015high}
\bibinfo{author}{Cheng, Y.s.}, \bibinfo{author}{Loo, B.P.},
  \bibinfo{author}{Vickerman, R.}, \bibinfo{year}{2015}.
\newblock \bibinfo{title}{{High-speed rail networks, economic integration and
  regional specialisation in China and Europe}}.
\newblock \bibinfo{journal}{Travel Behaviour and Society} \bibinfo{volume}{2},
  \bibinfo{pages}{1--14}.
\bibitem[{Cui and Li(2019)}]{cui2019high}
\bibinfo{author}{Cui, C.}, \bibinfo{author}{Li, L.S.Z.}, \bibinfo{year}{2019}.
\newblock \bibinfo{title}{{High-speed rail and inventory reduction: Firm-level
  evidence from China}}.
\newblock \bibinfo{journal}{Applied Economics} \bibinfo{volume}{51},
  \bibinfo{pages}{2715--2730}.
\bibitem[{Deng et~al.(2019)Deng, Wang, Yang and Yang}]{deng2019shrinking}
\bibinfo{author}{Deng, T.}, \bibinfo{author}{Wang, D.}, \bibinfo{author}{Yang,
  Y.}, \bibinfo{author}{Yang, H.}, \bibinfo{year}{2019}.
\newblock \bibinfo{title}{{Shrinking cities in growing China: Did high speed
  rail further aggravate urban shrinkage?}}
\newblock \bibinfo{journal}{Cities} \bibinfo{volume}{86},
  \bibinfo{pages}{210--219}.
\bibitem[{Diao(2018)}]{diao2018does}
\bibinfo{author}{Diao, M.}, \bibinfo{year}{2018}.
\newblock \bibinfo{title}{{Does growth follow the rail? The potential impact of
  high-speed rail on the economic geography of China}}.
\newblock \bibinfo{journal}{Transportation Research Part A: Policy and
  Practice} \bibinfo{volume}{113}, \bibinfo{pages}{279--290}.
\bibitem[{Dong et~al.(2021)Dong, Du, Kahn, Ratti and Zheng}]{dong2021ghost}
\bibinfo{author}{Dong, L.}, \bibinfo{author}{Du, R.}, \bibinfo{author}{Kahn,
  M.}, \bibinfo{author}{Ratti, C.}, \bibinfo{author}{Zheng, S.},
  \bibinfo{year}{2021}.
\newblock \bibinfo{title}{{"Ghost cities" versus boom towns: Do China's
  high-speed rail new towns thrive?}}
\newblock \bibinfo{journal}{Regional Science and Urban Economics} ,
  \bibinfo{pages}{103682}.
\bibitem[{Dong(2018)}]{dong2018highspeed}
\bibinfo{author}{Dong, X.}, \bibinfo{year}{2018}.
\newblock \bibinfo{title}{{High-speed railway and urban sectoral employment in
  China}}.
\newblock \bibinfo{journal}{Transportation Research Part A: Policy and
  Practice} \bibinfo{volume}{116}, \bibinfo{pages}{603--621}.
\bibitem[{Dong et~al.(2020)Dong, Zheng and Kahn}]{dong2020role}
\bibinfo{author}{Dong, X.}, \bibinfo{author}{Zheng, S.}, \bibinfo{author}{Kahn,
  M.E.}, \bibinfo{year}{2020}.
\newblock \bibinfo{title}{The role of transportation speed in facilitating high
  skilled teamwork across cities}.
\newblock \bibinfo{journal}{Journal of Urban Economics} \bibinfo{volume}{115},
  \bibinfo{pages}{103212}.
\bibitem[{Duan et~al.(2020)Duan, Sun and Zheng}]{duan2020transportation}
\bibinfo{author}{Duan, L.}, \bibinfo{author}{Sun, W.}, \bibinfo{author}{Zheng,
  S.}, \bibinfo{year}{2020}.
\newblock \bibinfo{title}{{Transportation network and venture capital mobility:
  An analysis of air travel and high-speed rail in China}}.
\newblock \bibinfo{journal}{Journal of Transport Geography}
  \bibinfo{volume}{88}, \bibinfo{pages}{102852}.
\bibitem[{Faber(2014)}]{faber2014trade}
\bibinfo{author}{Faber, B.}, \bibinfo{year}{2014}.
\newblock \bibinfo{title}{{Trade integration, market size, and
  industrialization: Evidence from China's National Trunk Highway System}}.
\newblock \bibinfo{journal}{Review of Economic Studies} \bibinfo{volume}{81},
  \bibinfo{pages}{1046--1070}.
\bibitem[{Fan(2005)}]{fan2005modeling}
\bibinfo{author}{Fan, C.C.}, \bibinfo{year}{2005}.
\newblock \bibinfo{title}{{Modeling interprovincial migration in China,
  1985-2000}}.
\newblock \bibinfo{journal}{Eurasian Geography and Economics}
  \bibinfo{volume}{46}, \bibinfo{pages}{165--184}.
\bibitem[{Fan et~al.(2019)Fan, Li, Zhang, Luo and Ma}]{fan2019connectivity}
\bibinfo{author}{Fan, J.}, \bibinfo{author}{Li, Y.}, \bibinfo{author}{Zhang,
  Y.}, \bibinfo{author}{Luo, X.}, \bibinfo{author}{Ma, C.},
  \bibinfo{year}{2019}.
\newblock \bibinfo{title}{Connectivity and accessibility of the railway network
  in {China: Guidance for spatial balanced} development}.
\newblock \bibinfo{journal}{Sustainability} \bibinfo{volume}{11},
  \bibinfo{pages}{7099}.
\bibitem[{Fang and Dewen(2003)}]{fang2003migration}
\bibinfo{author}{Fang, C.}, \bibinfo{author}{Dewen, W.}, \bibinfo{year}{2003}.
\newblock \bibinfo{title}{{Migration as marketidsszation: What can we learn
  from China's 2000 census data?}}
\newblock \bibinfo{journal}{China Review} \bibinfo{volume}{3},
  \bibinfo{pages}{73--93}.
\bibitem[{Gao et~al.(2021)Gao, Jun, Pentland, Zhou and
  Hidalgo}]{gao2021spillovers}
\bibinfo{author}{Gao, J.}, \bibinfo{author}{Jun, B.},
  \bibinfo{author}{Pentland, A.}, \bibinfo{author}{Zhou, T.},
  \bibinfo{author}{Hidalgo, C.A.}, \bibinfo{year}{2021}.
\newblock \bibinfo{title}{Spillovers across industries and regions in {China's}
  regional economic diversification}.
\newblock \bibinfo{journal}{Regional Studies} \bibinfo{volume}{55},
  \bibinfo{pages}{1311--1326}.
\bibitem[{Gao et~al.(2019a)Gao, Zhang and Zhou}]{gao2019computational}
\bibinfo{author}{Gao, J.}, \bibinfo{author}{Zhang, Y.C.},
  \bibinfo{author}{Zhou, T.}, \bibinfo{year}{2019}a.
\newblock \bibinfo{title}{Computational socioeconomics}.
\newblock \bibinfo{journal}{Physics Reports} \bibinfo{volume}{817},
  \bibinfo{pages}{1--104}.
\bibitem[{Gao et~al.(2020)Gao, Song, Sun and Zang}]{gao2020does}
\bibinfo{author}{Gao, Y.}, \bibinfo{author}{Song, S.}, \bibinfo{author}{Sun,
  J.}, \bibinfo{author}{Zang, L.}, \bibinfo{year}{2020}.
\newblock \bibinfo{title}{{Does high-speed rail connection really promote local
  economy? Evidence from China's Yangtze River Delta}}.
\newblock \bibinfo{journal}{Review of Development Economics}
  \bibinfo{volume}{24}, \bibinfo{pages}{316--338}.
\bibitem[{Gao et~al.(2019b)Gao, Su and Wang}]{gao2019does}
\bibinfo{author}{Gao, Y.}, \bibinfo{author}{Su, W.}, \bibinfo{author}{Wang,
  K.}, \bibinfo{year}{2019}b.
\newblock \bibinfo{title}{{Does high-speed rail boost tourism growth? New
  evidence from China}}.
\newblock \bibinfo{journal}{Tourism Management} \bibinfo{volume}{72},
  \bibinfo{pages}{220--231}.
\bibitem[{Guirao et~al.(2018)Guirao, Campa and Casado-Sanz}]{guirao2018labour}
\bibinfo{author}{Guirao, B.}, \bibinfo{author}{Campa, J.L.},
  \bibinfo{author}{Casado-Sanz, N.}, \bibinfo{year}{2018}.
\newblock \bibinfo{title}{{Labour mobility between cities and metropolitan
  integration: The role of high speed rail commuting in Spain}}.
\newblock \bibinfo{journal}{Cities} \bibinfo{volume}{78},
  \bibinfo{pages}{140--154}.
\bibitem[{Helpman and Krugman(1985)}]{helpman1985market}
\bibinfo{author}{Helpman, E.}, \bibinfo{author}{Krugman, P.R.},
  \bibinfo{year}{1985}.
\newblock \bibinfo{title}{{Market structure and foreign trade: Increasing
  returns, imperfect competition, and the international economy}}.
\newblock \bibinfo{publisher}{Cambridge, MA: MIT Press}.
\bibitem[{Heuermann and Schmieder(2019)}]{heuermann2019effect}
\bibinfo{author}{Heuermann, D.F.}, \bibinfo{author}{Schmieder, J.F.},
  \bibinfo{year}{2019}.
\newblock \bibinfo{title}{{The effect of infrastructure on worker mobility:
  Evidence from high-speed rail expansion in Germany}}.
\newblock \bibinfo{journal}{Journal of Economic Geography}
  \bibinfo{volume}{19}, \bibinfo{pages}{335--372}.
\bibitem[{Huang et~al.(2018)Huang, Levinson, Wang, Zhou and
  Wang}]{huang2018tracking}
\bibinfo{author}{Huang, J.}, \bibinfo{author}{Levinson, D.},
  \bibinfo{author}{Wang, J.}, \bibinfo{author}{Zhou, J.},
  \bibinfo{author}{Wang, Z.j.}, \bibinfo{year}{2018}.
\newblock \bibinfo{title}{Tracking job and housing dynamics with smartcard
  data}.
\newblock \bibinfo{journal}{Proceedings of the National Academy of Sciences}
  \bibinfo{volume}{115}, \bibinfo{pages}{12710--12715}.
\bibitem[{Huang and Xu(2021)}]{huang2021spatial}
\bibinfo{author}{Huang, Y.}, \bibinfo{author}{Xu, W.}, \bibinfo{year}{2021}.
\newblock \bibinfo{title}{{Spatial and temporal heterogeneity of the impact of
  high-speed railway on urban economy: Empirical study of Chinese cities}}.
\newblock \bibinfo{journal}{Journal of Transport Geography}
  \bibinfo{volume}{91}, \bibinfo{pages}{102972}.
\bibitem[{Jia et~al.(2017)Jia, Zhou and Qin}]{jia2017no}
\bibinfo{author}{Jia, S.}, \bibinfo{author}{Zhou, C.}, \bibinfo{author}{Qin,
  C.}, \bibinfo{year}{2017}.
\newblock \bibinfo{title}{{No difference in effect of high-speed rail on
  regional economic growth based on match effect perspective?}}
\newblock \bibinfo{journal}{Transportation Research Part A: Policy and
  Practice} \bibinfo{volume}{106}, \bibinfo{pages}{144--157}.
\bibitem[{Jiao et~al.(2017)Jiao, Wang and Jin}]{jiao2017impacts}
\bibinfo{author}{Jiao, J.}, \bibinfo{author}{Wang, J.}, \bibinfo{author}{Jin,
  F.}, \bibinfo{year}{2017}.
\newblock \bibinfo{title}{{Impacts of high-speed rail lines on the city network
  in China}}.
\newblock \bibinfo{journal}{Journal of Transport Geography}
  \bibinfo{volume}{60}, \bibinfo{pages}{257--266}.
\bibitem[{Jiao et~al.(2014)Jiao, Wang, Jin and Dunford}]{jiao2014impacts}
\bibinfo{author}{Jiao, J.}, \bibinfo{author}{Wang, J.}, \bibinfo{author}{Jin,
  F.}, \bibinfo{author}{Dunford, M.}, \bibinfo{year}{2014}.
\newblock \bibinfo{title}{{Impacts on accessibility of China’s present and
  future HSR network}}.
\newblock \bibinfo{journal}{Journal of Transport Geography}
  \bibinfo{volume}{40}, \bibinfo{pages}{123--132}.
\bibitem[{Jiao et~al.(2020)Jiao, Wang, Zhang, Jin and Liu}]{jiao2020roles}
\bibinfo{author}{Jiao, J.}, \bibinfo{author}{Wang, J.}, \bibinfo{author}{Zhang,
  F.}, \bibinfo{author}{Jin, F.}, \bibinfo{author}{Liu, W.},
  \bibinfo{year}{2020}.
\newblock \bibinfo{title}{{Roles of accessibility, connectivity and spatial
  interdependence in realizing the economic impact of high-speed rail: Evidence
  from China}}.
\newblock \bibinfo{journal}{Transport Policy} \bibinfo{volume}{91},
  \bibinfo{pages}{1--15}.
\bibitem[{Ke et~al.(2017)Ke, Chen, Hong and Hsiao}]{ke2017china}
\bibinfo{author}{Ke, X.}, \bibinfo{author}{Chen, H.}, \bibinfo{author}{Hong,
  Y.}, \bibinfo{author}{Hsiao, C.}, \bibinfo{year}{2017}.
\newblock \bibinfo{title}{{Do China's high-speed-rail projects promote local
  economy?—New evidence from a panel data approach}}.
\newblock \bibinfo{journal}{China Economic Review} \bibinfo{volume}{44},
  \bibinfo{pages}{203--226}.
\bibitem[{Kim and Sultana(2015)}]{kim2015impacts}
\bibinfo{author}{Kim, H.}, \bibinfo{author}{Sultana, S.}, \bibinfo{year}{2015}.
\newblock \bibinfo{title}{{The impacts of high-speed rail extensions on
  accessibility and spatial equity changes in South Korea from 2004 to 2018}}.
\newblock \bibinfo{journal}{Journal of Transport Geography}
  \bibinfo{volume}{45}, \bibinfo{pages}{48--61}.
\bibitem[{Krugman(1991)}]{krugman1991increasing}
\bibinfo{author}{Krugman, P.}, \bibinfo{year}{1991}.
\newblock \bibinfo{title}{Increasing returns and economic geography}.
\newblock \bibinfo{journal}{Journal of Political Economy} \bibinfo{volume}{99},
  \bibinfo{pages}{483--499}.
\bibitem[{Lawrence et~al.(2019)Lawrence, Bullock and Liu}]{lawrence2019china}
\bibinfo{author}{Lawrence, M.}, \bibinfo{author}{Bullock, R.},
  \bibinfo{author}{Liu, Z.}, \bibinfo{year}{2019}.
\newblock \bibinfo{title}{China's high-speed rail development}.
\newblock \bibinfo{publisher}{Washington, USA: World Bank}.
\bibitem[{Lee(1966)}]{lee1966theory}
\bibinfo{author}{Lee, E.S.}, \bibinfo{year}{1966}.
\newblock \bibinfo{title}{A theory of migration}.
\newblock \bibinfo{journal}{Demography} \bibinfo{volume}{3},
  \bibinfo{pages}{47--57}.
\bibitem[{Li et~al.(2020a)Li, Gao, Liang, Kang, Prestby, Gao and
  Xiao}]{li2020estimation}
\bibinfo{author}{Li, B.}, \bibinfo{author}{Gao, S.}, \bibinfo{author}{Liang,
  Y.}, \bibinfo{author}{Kang, Y.}, \bibinfo{author}{Prestby, T.},
  \bibinfo{author}{Gao, Y.}, \bibinfo{author}{Xiao, R.}, \bibinfo{year}{2020}a.
\newblock \bibinfo{title}{Estimation of regional economic development indicator
  from transportation network analytics}.
\newblock \bibinfo{journal}{Scientific Reports} \bibinfo{volume}{10},
  \bibinfo{pages}{2647}.
\bibitem[{Li et~al.(2020b)Li, Wu and Zhao}]{li2020economic}
\bibinfo{author}{Li, X.}, \bibinfo{author}{Wu, Z.}, \bibinfo{author}{Zhao, X.},
  \bibinfo{year}{2020}b.
\newblock \bibinfo{title}{{Economic effect and its disparity of high speed rail
  in China: A study of mechanism based on synthesis control method}}.
\newblock \bibinfo{journal}{Transport Policy} \bibinfo{volume}{99},
  \bibinfo{pages}{262--274}.
\bibitem[{Li et~al.(2013)Li, Liu and Tang}]{li2014analysis}
\bibinfo{author}{Li, Y.}, \bibinfo{author}{Liu, H.}, \bibinfo{author}{Tang,
  Q.}, \bibinfo{year}{2013}.
\newblock \bibinfo{title}{Analysis of determinants on {China's} interprovincial
  migration during 1985-2005}.
\newblock \bibinfo{journal}{Advanced Materials Research}
  \bibinfo{volume}{869--870}, \bibinfo{pages}{1096--1105}.
\bibitem[{Liang et~al.(2020)Liang, Zhou, Li, Zhou, Sun and
  Zeng}]{liang2020effectiveness}
\bibinfo{author}{Liang, Y.}, \bibinfo{author}{Zhou, K.}, \bibinfo{author}{Li,
  X.}, \bibinfo{author}{Zhou, Z.}, \bibinfo{author}{Sun, W.},
  \bibinfo{author}{Zeng, J.}, \bibinfo{year}{2020}.
\newblock \bibinfo{title}{Effectiveness of high-speed railway on regional
  economic growth for less developed areas}.
\newblock \bibinfo{journal}{Journal of Transport Geography}
  \bibinfo{volume}{82}, \bibinfo{pages}{102621}.
\bibitem[{Lin(2017)}]{lin2017travel}
\bibinfo{author}{Lin, Y.}, \bibinfo{year}{2017}.
\newblock \bibinfo{title}{{Travel costs and urban specialization patterns:
  Evidence from China's high speed railway system}}.
\newblock \bibinfo{journal}{Journal of Urban Economics} \bibinfo{volume}{98},
  \bibinfo{pages}{98--123}.
\bibitem[{Lin et~al.(2019)Lin, Qin, Sulaeman, Yan and
  Zhang}]{lin2019facilitating}
\bibinfo{author}{Lin, Y.}, \bibinfo{author}{Qin, Y.},
  \bibinfo{author}{Sulaeman, J.}, \bibinfo{author}{Yan, J.},
  \bibinfo{author}{Zhang, J.}, \bibinfo{year}{2019}.
\newblock \bibinfo{title}{Expanding footprints: {The} impact of passenger
  transportation on corporate locations}.
\newblock \bibinfo{note}{Available at SSRN: https://ssrn.com/abstract=3418227}.
\bibitem[{Lin et~al.(2021a)Lin, Qin, Wu and Xu}]{lingreenhouse}
\bibinfo{author}{Lin, Y.}, \bibinfo{author}{Qin, Y.}, \bibinfo{author}{Wu, J.},
  \bibinfo{author}{Xu, M.}, \bibinfo{year}{2021}a.
\newblock \bibinfo{title}{Impact of high-speed rail on road traffic and
  greenhouse gas emissions}.
\newblock \bibinfo{journal}{Nature Climate Change} \bibinfo{volume}{11},
  \bibinfo{pages}{952--957}.
\bibitem[{Lin et~al.(2021b)Lin, Qin and Xie}]{lintechnology}
\bibinfo{author}{Lin, Y.}, \bibinfo{author}{Qin, Y.}, \bibinfo{author}{Xie,
  Z.}, \bibinfo{year}{2021}b.
\newblock \bibinfo{title}{{Does foreign technology transfer spur domestic
  innovation? Evidence from the high-speed rail sector in China}}.
\newblock \bibinfo{journal}{Journal of Comparative Economics}
  \bibinfo{volume}{49}, \bibinfo{pages}{212--229}.
\bibitem[{Liu and Shen(2014)}]{liu2014spatial}
\bibinfo{author}{Liu, Y.}, \bibinfo{author}{Shen, J.}, \bibinfo{year}{2014}.
\newblock \bibinfo{title}{{Spatial patterns and determinants of skilled
  internal migration in China, 2000--2005}}.
\newblock \bibinfo{journal}{Papers in Regional Science} \bibinfo{volume}{93},
  \bibinfo{pages}{749--771}.
\bibitem[{Long et~al.(2018)Long, Zheng and Song}]{long2018high}
\bibinfo{author}{Long, F.}, \bibinfo{author}{Zheng, L.}, \bibinfo{author}{Song,
  Z.}, \bibinfo{year}{2018}.
\newblock \bibinfo{title}{{High-speed rail and urban expansion: An empirical
  study using a time series of nighttime light satellite data in China}}.
\newblock \bibinfo{journal}{Journal of Transport Geography}
  \bibinfo{volume}{72}, \bibinfo{pages}{106--118}.
\bibitem[{Lowry(1966)}]{lowry1966migration}
\bibinfo{author}{Lowry, I.S.}, \bibinfo{year}{1966}.
\newblock \bibinfo{title}{{Migration and metropolitan growth: Two analytical
  models}}.
\newblock \bibinfo{publisher}{San Francisco, USA: Chandler Publishing Company}.
\bibitem[{Muth(1969)}]{muth1969cities}
\bibinfo{author}{Muth, R.F.}, \bibinfo{year}{1969}.
\newblock \bibinfo{title}{{Cities and housing: The spatial patterns of urban
  residential land use}}.
\newblock \bibinfo{publisher}{Chicago, USA: University of Chicago Press.}
\bibitem[{Pagliara et~al.(2017)Pagliara, Mauriello and
  Garofalo}]{pagliara2017exploring}
\bibinfo{author}{Pagliara, F.}, \bibinfo{author}{Mauriello, F.},
  \bibinfo{author}{Garofalo, A.}, \bibinfo{year}{2017}.
\newblock \bibinfo{title}{{Exploring the interdependences between High Speed
  Rail systems and tourism: Some evidence from Italy}}.
\newblock \bibinfo{journal}{Transportation Research Part A: Policy and
  Practice} \bibinfo{volume}{106}, \bibinfo{pages}{300--308}.
\bibitem[{Pan et~al.(2020)Pan, Cong, Zhang and Zhang}]{pan2020high}
\bibinfo{author}{Pan, H.}, \bibinfo{author}{Cong, C.}, \bibinfo{author}{Zhang,
  X.}, \bibinfo{author}{Zhang, Y.}, \bibinfo{year}{2020}.
\newblock \bibinfo{title}{How do high-speed rail projects affect the
  agglomeration in cities and regions?}
\newblock \bibinfo{journal}{Transportation Research Part D: Transport and
  Environment} \bibinfo{volume}{88}, \bibinfo{pages}{102561}.
\bibitem[{Qin(2017)}]{qin2017no}
\bibinfo{author}{Qin, Y.}, \bibinfo{year}{2017}.
\newblock \bibinfo{title}{{‘No county left behind?'The distributional impact
  of high-speed rail upgrades in China}}.
\newblock \bibinfo{journal}{Journal of Economic Geography}
  \bibinfo{volume}{17}, \bibinfo{pages}{489--520}.
\bibitem[{Ravenstein(1889)}]{ravenstein1889law}
\bibinfo{author}{Ravenstein, E.G.}, \bibinfo{year}{1889}.
\newblock \bibinfo{title}{The laws of migration}.
\newblock \bibinfo{journal}{Journal of the Royal Statistical Society}
  \bibinfo{volume}{52}, \bibinfo{pages}{241--305}.
\bibitem[{Sasaki et~al.(1997)Sasaki, Ohashi and Ando}]{sasaki1997high}
\bibinfo{author}{Sasaki, K.}, \bibinfo{author}{Ohashi, T.},
  \bibinfo{author}{Ando, A.}, \bibinfo{year}{1997}.
\newblock \bibinfo{title}{{High-speed rail transit impact on regional systems:
  Does the Shinkansen contribute to dispersion?}}
\newblock \bibinfo{journal}{Annals of Regional Science} \bibinfo{volume}{31},
  \bibinfo{pages}{77--98}.
\bibitem[{Shao et~al.(2017)Shao, Tian and Yang}]{shao2017high}
\bibinfo{author}{Shao, S.}, \bibinfo{author}{Tian, Z.}, \bibinfo{author}{Yang,
  L.}, \bibinfo{year}{2017}.
\newblock \bibinfo{title}{{High speed rail and urban service industry
  agglomeration: Evidence from China's Yangtze River Delta region}}.
\newblock \bibinfo{journal}{Journal of Transport Geography}
  \bibinfo{volume}{64}, \bibinfo{pages}{174--183}.
\bibitem[{Shen(2012)}]{shen2012changing}
\bibinfo{author}{Shen, J.}, \bibinfo{year}{2012}.
\newblock \bibinfo{title}{{Changing patterns and determinants of
  interprovincial migration in China 1985--2000}}.
\newblock \bibinfo{journal}{Population, Space and Place} \bibinfo{volume}{18},
  \bibinfo{pages}{384--402}.
\bibitem[{Sjaastad(1962)}]{sjaastad1962costs}
\bibinfo{author}{Sjaastad, L.A.}, \bibinfo{year}{1962}.
\newblock \bibinfo{title}{The costs and returns of human migration}.
\newblock \bibinfo{journal}{Journal of Political Economy} \bibinfo{volume}{70},
  \bibinfo{pages}{80--93}.
\bibitem[{Tian et~al.(2021)Tian, Li, Ye, Zhao and Meng}]{tian2021impact}
\bibinfo{author}{Tian, M.}, \bibinfo{author}{Li, T.}, \bibinfo{author}{Ye, X.},
  \bibinfo{author}{Zhao, H.}, \bibinfo{author}{Meng, X.}, \bibinfo{year}{2021}.
\newblock \bibinfo{title}{The impact of high-speed rail on service industry
  agglomeration in peripheral cities}.
\newblock \bibinfo{journal}{Transportation Research Part D: Transport and
  Environment} \bibinfo{volume}{93}, \bibinfo{pages}{102745}.
\bibitem[{Vickerman and Ulied(2006)}]{vickerman2006indirect}
\bibinfo{author}{Vickerman, R.}, \bibinfo{author}{Ulied, A.},
  \bibinfo{year}{2006}.
\newblock \bibinfo{title}{Indirect and wider economic impacts of high speed
  rail}.
\newblock \bibinfo{journal}{Economic Analysis of High Speed Rail in Europe} ,
  \bibinfo{pages}{89--118}.
\bibitem[{Wang et~al.(2020a)Wang, Liu and Zhang}]{wang2020exploring}
\bibinfo{author}{Wang, C.}, \bibinfo{author}{Liu, H.}, \bibinfo{author}{Zhang,
  M.}, \bibinfo{year}{2020}a.
\newblock \bibinfo{title}{{Exploring the mechanism of border effect on urban
  land expansion: A case study of Beijing-Tianjin-Hebei region in China}}.
\newblock \bibinfo{journal}{Land Use Policy} \bibinfo{volume}{92},
  \bibinfo{pages}{104424}.
\bibitem[{Wang et~al.(2019a)Wang, Wei, Liu, He and Gao}]{wang2019impact}
\bibinfo{author}{Wang, F.}, \bibinfo{author}{Wei, X.}, \bibinfo{author}{Liu,
  J.}, \bibinfo{author}{He, L.}, \bibinfo{author}{Gao, M.},
  \bibinfo{year}{2019}a.
\newblock \bibinfo{title}{{Impact of high-speed rail on population mobility and
  urbanisation: A case study on Yangtze River Delta urban agglomeration,
  China}}.
\newblock \bibinfo{journal}{Transportation Research Part A: Policy and
  Practice} \bibinfo{volume}{127}, \bibinfo{pages}{99--114}.
\bibitem[{Wang et~al.(2020b)Wang, Xue, Chang and Xie}]{wang2020macroeconomic}
\bibinfo{author}{Wang, L.}, \bibinfo{author}{Xue, Y.}, \bibinfo{author}{Chang,
  M.}, \bibinfo{author}{Xie, C.}, \bibinfo{year}{2020}b.
\newblock \bibinfo{title}{{Macroeconomic determinants of high-tech migration in
  China: The case of Yangtze River Delta Urban Agglomeration}}.
\newblock \bibinfo{journal}{Cities} \bibinfo{volume}{107},
  \bibinfo{pages}{102888}.
\bibitem[{Wang et~al.(2019b)Wang, Dong, Liu, Huang and Liu}]{wang2019migration}
\bibinfo{author}{Wang, Y.}, \bibinfo{author}{Dong, L.}, \bibinfo{author}{Liu,
  Y.}, \bibinfo{author}{Huang, Z.}, \bibinfo{author}{Liu, Y.},
  \bibinfo{year}{2019}b.
\newblock \bibinfo{title}{Migration patterns in {China} extracted from mobile
  positioning data}.
\newblock \bibinfo{journal}{Habitat International} \bibinfo{volume}{86},
  \bibinfo{pages}{71--80}.
\bibitem[{Wu et~al.(2013)Wu, Fang, Zhao and Chen}]{wukang159}
\bibinfo{author}{Wu, K.}, \bibinfo{author}{Fang, C.}, \bibinfo{author}{Zhao,
  M.}, \bibinfo{author}{Chen, C.}, \bibinfo{year}{2013}.
\newblock \bibinfo{title}{{The intercity space of flow influenced by high-speed
  rail: A case study for the rail transit passenger behavior between Beijing
  and Tianjin}}.
\newblock \bibinfo{journal}{Acta Geographica Sinica} \bibinfo{volume}{68},
  \bibinfo{pages}{159--174}.
\bibitem[{Xu and Sun(2021)}]{xu2020siphon}
\bibinfo{author}{Xu, Z.}, \bibinfo{author}{Sun, T.}, \bibinfo{year}{2021}.
\newblock \bibinfo{title}{{The Siphon effects of transportation infrastructure
  on internal migration: Evidence from China's HSR network}}.
\newblock \bibinfo{journal}{Applied Economics Letters} \bibinfo{volume}{28},
  \bibinfo{pages}{1066--1070}.
\bibitem[{Yang et~al.(2018)Yang, Guo, Li and Huang}]{yang2018study}
\bibinfo{author}{Yang, J.}, \bibinfo{author}{Guo, A.}, \bibinfo{author}{Li,
  X.}, \bibinfo{author}{Huang, T.}, \bibinfo{year}{2018}.
\newblock \bibinfo{title}{{Study of the impact of a high-speed railway opening
  on China's accessibility pattern and spatial equality}}.
\newblock \bibinfo{journal}{Sustainability} \bibinfo{volume}{10},
  \bibinfo{pages}{2943}.
\bibitem[{Yang et~al.(2019)Yang, Lin, Zhang and He}]{yang2019does}
\bibinfo{author}{Yang, X.}, \bibinfo{author}{Lin, S.}, \bibinfo{author}{Zhang,
  J.}, \bibinfo{author}{He, M.}, \bibinfo{year}{2019}.
\newblock \bibinfo{title}{Does high-speed rail promote enterprises
  productivity?{ Evidence from China}}.
\newblock \bibinfo{journal}{Journal of Advanced Transportation}
  \bibinfo{volume}{2019}, \bibinfo{pages}{1279489}.
\bibitem[{Yin et~al.(2015)Yin, Bertolini and Duan}]{yin2015effects}
\bibinfo{author}{Yin, M.}, \bibinfo{author}{Bertolini, L.},
  \bibinfo{author}{Duan, J.}, \bibinfo{year}{2015}.
\newblock \bibinfo{title}{{The effects of the high-speed railway on urban
  development: International experience and potential implications for China}}.
\newblock \bibinfo{journal}{Progress in Planning} \bibinfo{volume}{98},
  \bibinfo{pages}{1--52}.
\bibitem[{Yu et~al.(2019)Yu, Lin, Tang and Zhong}]{yu2019highspeed}
\bibinfo{author}{Yu, F.}, \bibinfo{author}{Lin, F.}, \bibinfo{author}{Tang,
  Y.}, \bibinfo{author}{Zhong, C.}, \bibinfo{year}{2019}.
\newblock \bibinfo{title}{{High-speed railway to success? The effects of
  high-speed rail connection on regional economic development in China}}.
\newblock \bibinfo{journal}{Journal of Regional Science} \bibinfo{volume}{59},
  \bibinfo{pages}{723--742}.
\bibitem[{Zhang et~al.(2018)Zhang, Yu, Zhong and Lin}]{zhang2018high}
\bibinfo{author}{Zhang, M.}, \bibinfo{author}{Yu, F.}, \bibinfo{author}{Zhong,
  C.}, \bibinfo{author}{Lin, F.}, \bibinfo{year}{2018}.
\newblock \bibinfo{title}{High-speed railways, market access and enterprises'
  productivity}.
\newblock \bibinfo{journal}{China Industrial Economics} \bibinfo{volume}{5},
  \bibinfo{pages}{137--156}.
\bibitem[{Zhou et~al.(2018)Zhou, Yang and Li}]{zhou2018implications}
\bibinfo{author}{Zhou, J.}, \bibinfo{author}{Yang, L.}, \bibinfo{author}{Li,
  L.}, \bibinfo{year}{2018}.
\newblock \bibinfo{title}{{The implications of high-speed rail for Chinese
  cities: Connectivity and accessibility}}.
\newblock \bibinfo{journal}{Transportation Research Part A: Policy and
  Practice} \bibinfo{volume}{116}, \bibinfo{pages}{308--326}.
\bibitem[{Zhou(2021)}]{zhou2021representative}
\bibinfo{author}{Zhou, T.}, \bibinfo{year}{2021}.
\newblock \bibinfo{title}{Representative methods of computational
  socioeconomics}.
\newblock \bibinfo{journal}{Journal of Physics: Complexity}
  \bibinfo{volume}{2}, \bibinfo{pages}{031002}.
\bibitem[{Zou et~al.(2021)Zou, Chen and Xiong}]{zou2019high}
\bibinfo{author}{Zou, W.}, \bibinfo{author}{Chen, L.}, \bibinfo{author}{Xiong,
  J.}, \bibinfo{year}{2021}.
\newblock \bibinfo{title}{{High-speed railway, market access and economic
  growth}}.
\newblock \bibinfo{journal}{International Review of Economics \& Finance}
  \bibinfo{volume}{76}, \bibinfo{pages}{1282--1304}.

\end{thebibliography}

\end{document}